\RequirePackage{fix-cm}
\documentclass[twocolumn]{svjour3custom}
\pdfoutput=1

\usepackage{float,algorithm,algorithmic,latexsym,enumerate}
\usepackage{graphicx}
\usepackage{tabularx}
\usepackage{alltt}
\usepackage[justification=centering]{caption}
\usepackage{multirow}
\usepackage{xcolor}
\usepackage{balance}
\usepackage{arydshln}
\usepackage{microtype}
\usepackage{amssymb}
\usepackage[colorlinks, citecolor=black, filecolor=black, 
linkcolor=black,urlcolor=blue]{hyperref}
\newfloat{algorithm}{t}{lop}

\hypersetup{colorlinks, citecolor=black, filecolor=black, linkcolor=black,urlcolor=blue}

\newcommand{\LOJoin}{\mathrel{\hbox{${_{\raise3pt\hbox{--}}^{\hbox{--}}}$}\hspace{-1.55mm}\Join}}
\newcommand{\ROJoin}{\Join\hspace{-1.55mm}\mathrel{\hbox{${~_{\raise3pt\hbox{--}}^{\hbox{--}}}$}}}
\newcommand{\FOJoin}{\mathrel{\hbox{${_{\raise3pt\hbox{--}}^{\hbox{--}}}$}\hspace{-1.5mm}\Join}\hspace{-1.5mm}\mathrel{\hbox{${~_{\raise3pt\hbox{--}}^{\hbox{--}}}$}}}

\newtheorem{exmp}{Example}

\floatstyle{boxed}
\newfloat{program}{thp}{lop}

\newcommand\blfootnote[1]{%
  \begingroup
  \renewcommand\thefootnote{}\footnote{#1}%
  \addtocounter{footnote}{-1}%
  \endgroup
}

\newfloat{cvcExample}{thp}{lop}
\newcommand{\newstuff}[1]{#1}

\newcommand{\reminder}[1]{}
\newcommand{\eat}[1]{}
\newcommand{\fullversion}[1]{}

\newcommand{\supplement}[1]{Appendix~\ref{#1}}

\newcommand{\smalltt}[1]{{\small{\texttt{#1}}}}

\newcommand{\query}[1]{\item \begin{small}\begin{alltt}#1\end{alltt}\end{small}}
\newcommand{\qindent}{\hspace*{0.75em}}

\newcommand{\likeop}{\textit{likeop}}
\newcommand{\relop}{\textit{relop}}

\begin{document}
 
 \pagestyle{plain}
 \title{Data Generation for Testing and Grading SQL Queries}
 
 \author{ \hspace{2mm}Bikash Chandra \and
        Bhupesh Chawda$~^{*}$  \and
        Biplab Kar$~^{\S}$ \and \\
        K. V. Maheshwara Reddy$~^{\#}$ \and 
        Shetal Shah \and
        S. Sudarshan \\ IIT Bombay \\
        {\{bikash, biplabkar11, kvmahesh12, shetals, sudarsha\}@cse.iitb.ac.in, bhchawda@in.irl.com}
}
 
\eat{
\institute{Bikash Chandra 
           \and Biplab Kar {$^\S$} 
	   \and K V Maheshwara Reddy {$^\#$}
	   \and Shetal Shah 
	   \and S Sudarshan
	   \and Bhupesh Chawda{$^*$}
	   \at \\IIT Bombay
	   \at \email{\{bikash, biplabkar11, kvmahesh12, shetals, \\sudarsha\}@cse.iitb.ac.in, bhchawda@in.irl.com}	   
	   \\{$^\#$} Currently working at SAP Labs, India
	   \\{$^\S$} Currently working at Oracle, India
	   \\{$^*$} Currently working at IBM IRL, India
}
}
\date{}

 \maketitle
 \blfootnote{
\hspace*{-0.82cm}$~^{\#}$~Currently working at SAP Labs, India\\$*$~Current working at IBM IRL, India\\
$~^{\S}$ Currently working at Oracle India Pvt. Ltd.
}
 \begin{abstract}

Correctness of SQL queries is usually tested by executing the queries on one or more datasets. 
Erroneous queries are often the results of small changes or mutations of the correct query. 
A mutation Q' of a query Q is killed by a dataset D if Q(D) $\neq$ Q'(D). 
Earlier work on the XData system showed how to
generate datasets that kill all mutations in a class of mutations that included join type and comparison operation
mutations.

In this paper, we extend the XData data generation techniques to handle a wider variety of SQL queries and a much larger class of mutations.
We have also built a system for grading SQL queries using the datasets generated by XData. 
We present a study of the effectiveness of the datasets generated by the extended XData
approach, using a variety of queries including queries submitted by students as part of
a database course. We show that the XData datasets outperform predefined datasets
as well as manual grading done earlier by teaching
assistants, while also avoiding the drudgery of manual correction. Thus, we believe that our techniques will be of 
great value to database course instructors and TAs, particularly to those of MOOCs.
It will also be valuable to database application developers and testers for testing SQL queries. 
\end{abstract}

 \section{Introduction}
\label{sec:intro}
Queries written in SQL are used in a variety of different applications. 
An important part of testing these applications is to test the correctness of SQL queries in these applications.
The queries are usually tested using multiple ad hoc test cases provided by the programmer or the tester. 
Queries are run against these test cases and tested by comparing the results with the intended one manually or by automated test cases.
However, this approach involves manual effort in terms of test case generation and also does not ensure whether all the relevant test cases have been covered or not.
Formal verification techniques involve comparing a specification with an implementation. 
However, since SQL queries are themselves specifications and do not contain the implementation, formal verification techniques cannot be applied for testing SQL queries.

A closely related problem is grading SQL queries written by students. 
Grading SQL queries is usually done by executing the query on small datasets and/or by reading the student query and comparing those with the correct query. 
Manually created datasets, as well as datasets created in a query independent manner, can be  incomplete and are likely to miss errors in queries.
Manual reading and comparing of queries is difficult, since students may write queries in a variety of different ways,
and is prone to errors as graders are likely to miss subtle mistakes. For example, when required to write the query $Q$ below:

\vspace{-4pt}\newstuff{
\begin{small}
\begin{alltt}
SELECT course.id, department.dept\_name  FROM 
course LEFT OUTER JOIN (SELECT * from department \\ WHERE department.budget > 70000) d USING (dept\_name);
\end{alltt}
\end{small}

\vspace{-4pt}
\noindent students often write the query $Q_s$:

\vspace{-4pt}
\begin{small}
\begin{alltt}
SELECT course.id, department.dept\_name FROM 
course LEFT OUTER JOIN department USING (dept\_name) \\ WHERE department.budget > 70000;
\end{alltt}
\end{small}} 
\noindent which looks sufficiently similar for a grader to miss the difference. 
These queries are not equivalent since they give different results on departments with
budget less than 70000.

\newstuff{Mutation testing is a well-known approach for checking
the adequacy of test cases for a program \cite{mutation:testing}}. Mutation testing
involves generating mutants of the original program by modifying the program in a controlled manner.
For SQL queries, we consider that a \textit{mutation} is a single (syntactically correct)
change of the original query; a \textit{mutant} is the result of one of more
mutations on the original query. 
A dataset \textit{kills} a mutant if the original query and the mutant
give different results on the dataset, allowing us to distinguish between the queries. 
A test suite consisting of multiple datasets kills a mutant if at least one of the datasets kills the mutant.

\vspace{1cm}
Consider the query:

\begin{small}
\begin{alltt}
SELECT dept\_name, COUNT(DISTINCT id) FROM
course LEFT OUTER JOIN takes
USING(course_id) GROUP BY dept\_name
\end{alltt}
\end{small}
One of the mutants obtained by mutating the join condition of the query is:

\begin{small}
\begin{alltt}
SELECT dept\_name, COUNT(DISTINCT id) FROM
course \textit{INNER JOIN} takes
USING(course\_id) GROUP BY dept\_name
\end{alltt}
\end{small}
Similarly by mutating the aggregation we get the following mutation:

\begin{small}
\begin{alltt}
SELECT dept\_name, \textit{COUNT(id)} FROM
course LEFT OUTER JOIN takes
USING(course_id) GROUP BY dept\_name
\end{alltt}
\end{small}

\newstuff{
In this paper, we address the problem of generating datasets that can catch commonly occurring errors 
in a large class of SQL queries. 
Queries with common errors can be thought of as  mutants of the original query.
Our goal is to generate (a relatively small number of) datasets so as to kill a 
wide variety of query mutations. } 
These datasets can be used in two distinct ways:
\begin{enumerate}[a)]
\item To check if a given query is what was intended, a tester manually examines 
the result of the query on each dataset, and checks if the result is what
was intended.
\item To check if a student query is correct, the results of the student query and a given correct query are
compared on each dataset. A difference on any dataset indicates that the student query is erroneous (We note that checking query equivalence is possible in limited special cases but is hard or undecidable in general\cite{ineq.equiv,conjuct.equiv,bag}).
\end{enumerate}

There has been increased interest in the recent years in test data generation for SQL queries including \cite{Tuya:2010,Riva:2010,SQE,QEX}; \cite{olstonCS09} addresses a similar problem in
the context of data-flow programs.  
Our earlier work on the XData system \cite{xdata:icde10,xdata:icde11} showed how to generate datasets that can distinguish the correct query from some class of query mutations, including join and comparison operator mutations. 
However, real life SQL queries have a variety of features and mutations 
that were not handled in \cite{xdata:icde10,xdata:icde11}.
(Related work is described in detail in Section~\ref{sec:relwork}.)
A few of the techniques described in this paper were sketched in a short workshop paper
\cite{xdata:dbtest13}, but details were not presented there.


\newstuff{
In Sections~\ref{sec:string} to \ref{sec:set} we describe techniques to handle different SQL query features. 
For each feature, we first discuss techniques to handle data generation for that feature, then describe 
mutations of these features, and finally present techniques to kill these mutations. In Section~\ref{sec:join:group:distinct} we describe techniques for killing new classes of mutations for query features
that were handled in our earlier work \cite{xdata:icde10,xdata:icde11}.

Each data generation technique is designed to handle specific query constructs or specific mutations of the query.
We combine these techniques to generate datasets for a complete query, with each dataset targeting a
specific type of mutation. One dataset is capable of killing one or more mutations. 
Specifically, we do not generate any mutants at all. Our goal is to generate datasets to kill mutations 
and not enumerate the possible mutants. Although the number of mutations may be very large, our approach generates
a small number of datasets that can kill a much larger number of mutations.  
}


The contributions of this paper are as follows. 
\begin{enumerate}
\item We discuss (in Section~\ref{sec:string}) how to generate test data and kill mutations for queries involving string predicates such as string
comparison and the LIKE predicate, using a string solver we have developed.
\item We support the NULL values and several mutations that may arise because of the presence of NULLs (Section~\ref{sec:nulls}).
\item For queries containing constraints on aggregated results, we describe (in Section~\ref{sec:constrainedagg}) a new algorithm to 
find the number of tuples that need to be generated for each relation to satisfy the aggregation constraints.
\item We support test data generation and mutation killing for a large class of nested subqueries (Section~\ref{sec:subquery}). 
\item We also support data generation and mutation killing for queries containing set operators (Section~\ref{sec:set}).
\item We extend the class of mutations considered to include missing or additional join conditions (Section~\ref{sec:joincond}), missing or additional group by attributes (Section~\ref{sec:group}), and  distinct clause mutations (Section~\ref{sec:distinct}). 
\item The data types supported include floating point numbers, time and date values. The class of queries is extended to include insert, delete, update and parameterized queries as well as  view creation  statements (Section~\ref{sec:misc}). 
\item We describe (in Section~\ref{sec:tool}) techniques for grading student queries based on the datasets generated by XData. 
These techniques can be used for grading, as well as in a 
learning mode where it can give immediate feedback to students.  
\item In Section~\ref{sec:expt} we present performance results of our techniques. 
We generate test data for a number of queries involving constrained aggregation and subqueries on the University database \cite{dbconcepts2010} as well as queries of the TPC-H benchmark and show that the datasets generated by XData are able to kill most of the non-equivalent mutations. 
We also test the effectiveness of our grading tool by using as a benchmark
a set of assignments given as part of a database course at IIT Bombay.
We show that the datasets generated using our techniques
catch more errors than the University datasets,
provided with \cite{dbconcepts2010}, 
as well as manual grading by the TAs, on all the queries.  
\end{enumerate}

We believe the techniques presented in this paper will be of 
great value to database application developers and testers for testing real life SQL queries.
It will also be valuable to database course instructors and TAs by taking the drudgery out of grading
and allow SQL query assignments to be properly checked in MOOC setting, where manual grading is not feasible.

 \section{Background}
\label{sec:bg}
In our earlier work on XData \cite{xdata:icde11}, we presented techniques for
generating test data for killing SQL query mutants; we briefly outline that work below.  

\subsection{Approach to Data Generation}
Given an SQL query $Q$, XData\cite{xdata:icde11}  generates multiple datasets.
\newstuff{The first dataset is designed to generate non-empty datasets for $Q$, wherever feasible, which itself kills several mutations
that would generate an empty result on that dataset. }
Each of the remaining datasets is targeted to kill one or more mutations of the query; i.e. on each dataset the
given query returns a result that is different from those returned by each of the 
mutations targeted by that dataset. The number of possible mutations is very large,
but the number of datasets generated to kill these mutations is small.

\newstuff{
To generate a particular dataset, XData does the following:
\begin{enumerate}
 \item It generates a set of constraint variables,
where each tuple in the target dataset is represented by a tuple of constraint variables.
  \item It generates a set of constraints between these variables. For example, selection conditions, 
  join conditions, primary key and foreign key conditions are all mapped to constraints on these variables. 
  Different datasets are designed to catch different mutations; the exact set of constraints generated (as also 
  the set of constraints variables) is different for each dataset, as described shortly.
  \item It then invokes a constraint (SMT) solver \cite{smt}\footnote{A constraint solver takes as input a set of constraints and produces a result that satisfies the constraints.} to solve the constraints;
the solution given by the solver assigns values to each constraint variable, thereby
  defining a specific dataset. 
\end{enumerate}
}

In order to kill mutations, the goal of XData is to generate datasets that produce different results on the query and its mutation. 
To produce different results, constraints are added in a manner so as to ensure that the mutation in a node of a query tree is reflected above leading to different results for the query and its mutation.
For example consider the following query:

\eat{
\begin{exmp} 
\label{ex:bg}
\begin{small}
\begin{verbatim}

SELECT course_id, dept_name, budget 
FROM course INNER JOIN department USING (dept_name)
WHERE dept_budget > 70000
\end{verbatim}
\end{small}
\end{exmp}

This query has two predicates \texttt{course INNER JOIN department USING (dept\_name)} and \texttt{dept\_budget > 70000}. 
When generating datasets to kill the mutations of join predicates we need to ensure that $dept\_budget>$ $70000$ is satisfied for the tuple generated for the \texttt{department} table. 
In case $dept\_budget > 70000$ is not satisfied both the query and the mutant could give empty result. 
}

\begin{exmp} 
\label{ex:bg}
\begin{small}
\begin{verbatim}

SELECT course.course_id, COUNT(DISTINCT takes.id) 
FROM course INNER JOIN takes USING(course_id)
WHERE course.credits >= 6
\end{verbatim}
\end{small}
\end{exmp}

\newstuff{This query has two predicates \texttt{course INNER JOIN takes USING (course\_id)} and \texttt{course.credits >= 6}. 
When generating datasets to kill the mutations of join predicates we need to ensure that $course.credits >= 6$ is satisfied for the tuple generated for the \texttt{course} table. 
In case $course.credits >= 6$ is not satisfied, both the query and the mutant could give empty results. }

\subsection{Mutation Space and Datasets}
\label{sec:bg:mutation}

The mutation space
considered consisted of the following
\begin{enumerate}
\item \textit{Join Type Mutations}: 
A join type mutations involves replacing one of \{ INNER, LEFT OUTER, RIGHT OUTER \} JOIN with another. 
Consider the mutation from \texttt{department INNER JOIN course} to \texttt{department LEFT OUTER JOIN course}.
In order to kill this mutation, we need to ensure that there is a tuple in department relation that does not satisfy the join condition with any tuple in course relation. 
\newstuff{The INNER JOIN query would not output that tuple in the department relation while the LEFT OUTER JOIN would.}

In SQL, a join query can be specified in a join order independent fashion, 
with many equivalent join orders for a given query.  
Hence, the number of join type mutations across all these 
orders is exponential. From the join conditions specified in the query, XData forms equivalence classes
of $<$relation, attribute$>$ pairs such that elements in the same equivalence class need to be assigned the same value to meet (one or more) join conditions.
Using these equivalence classes, XData generates a {\em linear} number of datasets to 
kill join type mutations across all join orderings.  If a pair of relations involve multiple join conditions XData nullifies each join condition separately.

\item \textit{Selection Predicate Mutations}: For selection conditions XData considers mutations of the \newstuff{relational operator} where any occurrence of one of $\{=, <>, <, >, \leq, \geq\}$ is replaced by another.
For killing mutations for the selection condition $A_1$ \relop\ $A_2$, XData generates 3 datasets (1)   $A_1 > A_2$, (2)  $A_1 < A_2$, and (3)  $A_1 = A_2$.
These three datasets kill all non-equivalent mutations from one \relop\ to another \relop. 
These datasets also kill mutations because of missing selection conditions. 
 
\item \textit{Unconstrained Aggregation Mutation}: Aggregations at the root of the query tree are not constrained to satisfy any condition. 
The aggregation function can be mutated among MAX, MIN, SUM, AVG, COUNT and their DISTINCT versions. 
In order to kill these mutations, a dataset with three tuples is generated; two with the same value (non-zero) and another with a different value in the aggregate column. 
\end{enumerate}

\subsection{Constraint Generation}
\newstuff{We now describe our techniques for constraint generation. 
Our current implementation uses the CVC3 constraint solver \cite{CVC3}. 
  We are working on implementing the constraints in the SMT-LIB format \cite{smtlib} 
  so that we can potentially use several constraint solvers compatible with SMT-LIB.}

In CVC3, text attributes are modeled as enumerated types while
numeric attributes are modeled as subtypes of integers or
rationals. The data type declarations in CVC3 are as follows.
For each attribute of each relation, we specify
a set of acceptable values, taken from an input database,
as datatypes in CVC3.  
While the input database is not necessary  
for data generation, its use makes for improved readability and 
comprehension of the query results.
In case an input database is not specified we get the range from the data type of the corresponding column. 

A tuple type is created for each relation, where each element is a constraint variable
of the specified type.
A relation is represented as an array of constraint variables;
the size of the array has to be determined before solving the constraints, and constraints
have to be specified for each attribute of each tuple.

\newstuff{
Consider an input database which has 
{\small{\texttt{CS-101}, \texttt{BIO-301}, \texttt{CS-312} and \texttt{PHY-101}}} as  
\texttt{course\_id}, and \texttt{credits} is an integer constrained to be between
2 and 10. Then, this translates to the following 
the declarations in CVC3. 
}

\begin{small}
\begin{alltt}
DATATYPE
course_id = CS-101 | BIO-301 | CS-312 | PHY-101 END;
credits:TYPE = SUBTYPE (LAMBDA (x: INT):
\hspace*{2.25cm}x > 1 AND x < 11);
course_tuple_type:TYPE = [course_id,credits];
course: ARRAY INT OF course_tuple_type;
\end{alltt}
\end{small}

\newstuff{
Tuple attributes are referenced by position, not by name; thus,
\texttt{course[2].0} refers to the value of the first attribute, which is \texttt{course\_id},
of the second tuple in \texttt{course}.

To ensure a non-empty result for the query in Example~\ref{ex:bg}, we need a tuple in \texttt{course} 
which matches a tuple in \texttt{takes} on attribute \texttt{course\_id} and where the \texttt{course.credits >= 6}. This is done by creating a tuple for each of the relations and adding the following constraints:\\
{\texttt{\small
ASSERT course[1].0 = takes[1].1;\\
ASSERT course[1].1 >= 6;
}}

Primary key constraints are enforced by constraints that ensure 
that if two tuples match on the primary key, then the values of the
remaining attributes for those two tuples should also match. 
Foreign key constraints are enforced by adding extra tuples that satisfy the foreign key condition. 
Foreign key constraints for the foreign key from \texttt{takes.course\_id} to \texttt{course.course\_id} 
are specified as: }

\begin{small}
\begin{alltt}
ASSERT FORALL(i: takes_index):
EXISTS (j: course_index): takes[i].1 = course[j].0;
\end{alltt}
\end{small}
\newstuff{where \texttt{takes\_index} and \texttt{course\_index} give the index range 
for the \texttt{takes} and \texttt{course} arrays; \texttt{takes[i].1} 
stands for \texttt{dept\_name} of the \texttt{i}$^{th}$ tuple of \texttt{course}.
In our example
an extra tuple would be generated for \texttt{course} for each tuple in \texttt{takes}, 
although in this case the first tuple of \texttt{course} itself ensures the 
foreign key constraint is satisfied for the first tuple of \texttt{takes}.

The above constraints are given to CVC3 which generates satisfying values 
(assuming the constraints are satisfiable).

As explained earlier in this section, to kill a mutation of the inner join to right outer join, we
need a value in \texttt{course.course\_id} which does not match any value in \texttt{takes.course\_id}.
To do so we replace the earlier equality constraint \\
{\texttt{\small
ASSERT course[1].0 = takes[1].1;
}}
\\with:\\
\hspace*{0.5cm} {\texttt{\small
ASSERT NOT EXISTS(i:course\_index):  \\
\hspace*{1cm}  (course[i].0 = takes[1].1);
}}\\
and generate the required dataset using CVC3.
Datasets for killing other mutations are generated similarly.
}

\subsection{Disjunctions}
\label{sec:disjunct}
\newstuff{
Tuya et al. in \cite{Riva:2010} presented techniques 
for killing mutations in the presence of disjunctions. 

For killing a where clause mutation of a query, the mutation should be 
reflected as a change at the root of the query tree. 
Consider the $where$ clause $P_{1}$ $or$ $P_{2}$ , where $P_{1}$ and 
$P_{2}$ are conjuncts of selection conditions. 
If a condition in  $P_{1}$ is mutated, $P_{2}$ should be false so 
that the change in the condition of $P_{1}$ affects the output of 
the query. For example, let $P_{1}$ be $(a > 50$ AND $b = 40)$. 
If we mutate the first condition in $P_{1}$ to $a<50$ we need to 
ensure that  $b=40$ is satisfied while $P_{2}$ is not satisfied. 
If $P_{2}$ is satisfied there would be no change in the output of 
the query. 
Although not mentioned in \cite{Riva:2010}, the above technique not only kills
mutations of atomic selection conditions (such as comparisons) but also kills mutations of conjunction operations 
to disjunctions and vice versa.

The XData system has been extended to implement
the above technique for killing selection predicate mutations in the presence of disjunctions. 

}

 \section{Queries and Mutations Considered}
\label{sec:mutant:space}


\label{classofQueries}

\newstuff{
The class of queries considered by XData now includes 
\begin{enumerate}[a)]
 \item Single block queries with join/outer-join operations 
and predicates in the where-clause, and optionally aggregate operations, 
corresponding to select / project / join / outer-join qu\-eries 
in relational algebra, with aggregation operations.
\item Multi-block queries. Our current implementation can deal with subqueries up to a single level of nesting. 
\item Compound queries with set operators UNION(ALL), INTERSECT(ALL) and EXCEPT(ALL). 
\end{enumerate}

}

In this paper, we remove the following assumptions made in \cite{xdata:icde11}:
\begin{enumerate}[a)]
 \item SQL queries do not contain string comparison or string like operators such as \textit{like, ilike}, etc.
 \item Aggregations are only present at the top of the query tree, and hence they are not constrained.
 \item SQL queries are single block queries with no nested subqueries.
 \item NULL values are not allowed for attribute values. 
 \item Selection predicates are conjunctions of simple conditions of the form \textit{expr relop expr}.
\end{enumerate}

\newstuff{
XData now considers a large class of mutations  - join type mutations, comparison operator mutations, aggregation mutations, 
string mutations, NULL mutations, set operator mutations,
join condition mutations, group by attribute mutations and distinct mutations. Of these only join type mutations, comparison operator mutations and aggregation mutations were discussed previously in \cite{xdata:icde11}.

\eat{In this paper we consider string mutations, aggregation mutations with constraints on aggregations, mutations due to NULL values, subquery mutations, set operator mutations, join conditions mutations, group by attribute mutations and distinct mutation that were not considered earlier.}
}


We retain the following assumptions
\begin{enumerate}[a)]
\item The only database constraints are unique, primary key and foreign key constraints.
\item Queries do not include numeric functions or expressions other than simple arithmetic expressions. 
\item Join predicates are conjunctions of simple conditions.
\item No user defined functions are used.
\end{enumerate}

We only consider single mutations in a query when generating test datasets, since the space of mutants is much larger with multiple mutations.
It is possible that an
erroneous query may contain multiple mistakes; queries with multiple mutations are likely,
but not always guaranteed, to be killed by the datasets we generate.
\newstuff{Completeness guarantees for our data generation techniques are described in \supplement{sec:app:complete}.}

 \section{Data Generation for String Constraints} 
\label{sec:string}

SQL queries can have equality and inequality conditions on strings, and 
pattern matching conditions using the LIKE operator or its variants. \\
\noindent Consider the SQL query,

\begin{small}
\begin{alltt}
SELECT * from student WHERE name LIKE `Amol\%'
AND name LIKE `\%Pal' AND tot_cred > 30
\end{alltt}
\end{small}

In order to generate the first dataset that produces a  non-empty result for this query or to kill mutations of the condition \texttt{tot\_cred > 30},
we need to generate a tuple for which attribute name satisfies the LIKE conditions `Amol\%' and `\%Pal'. 
To generate such a value we need to solve the corresponding string constraints. 
For killing mutations of the LIKE operators also we need to solve similar string constraints. 

Since many constraint solvers, including CVC3, do not support string constraints, 
we solve the string constraints outside of the solver.
\newstuff{We describe the types of string constraints considered in Section~\ref{sec:string:type} and our approach to solving string constraints in Section~\ref{sec:string:solver}}
 We then discuss test data generation for killing mutations involving string operators in Section~\ref{sec:string:mutant}.
\newstuff{Note that for this to work; there should be no dependence 
between string and other constraints so that the string constraints 
can be solved independently of other constraints. 
For example, for constraints like $length(R.a)>R.b$, where $R.a$ is a string 
attribute and $R.b$ is an integer attribute, the condition on $R.a$ cannot
be solved independently of constraints on $R.b$ if there are other constraints 
on $R.a$ and $R.b$.
However, if an integrated  constraint solver this restriction does not apply.}


\subsection{Types of String Constraints Considered}
\label{sec:string:type}
For string comparisons, we consider the following class of 
string constraints: $S_1$ \relop\ constant,
and $S_1$ \relop\ $S_2$, where $S_1$ and $S_2$ are string variables, and
\relop\ operators are $=, <, \leq, >, \geq, <>$ 
and case-insensitive equality denoted by $\sim =$.
\eat{(For $<>$ and $\sim =$, our solver is incomplete if both operands 
are variables, and both variables have other constraints on them;
thus we assume that one of the operands is a constant or a variable with
no other constraint on it.)}
We support LIKE constraints of the form
$S$ \likeop\ pattern, where \likeop\ is one of \textit{LIKE, ILIKE} (case insensitive like), 
\textit{NOT LIKE} and \textit{NOT ILIKE}. 
We also support \textit{strlen}($S$) \relop\  constant
where \relop\ is one of $=, <, \leq, >, \geq$ or $<>$.
We do not support constraints of the form $S_1$ \likeop\ $S_2$, where both $S_1$ and $S_2$ are variables.

We support the string functions \textit{upper} and \textit{lower} 
in queries where these functions can be rewritten
using one of the operators described above; for example \textit{upper}($S$) = \textquoteleft ABC\textquoteright \ can be rewritten
as $S \sim =$  \textquoteleft ABC\textquoteright, and similarly \textit{upper}($S$) \textit{LIKE pattern} can be replaced
by $S$ \textit{ILIKE pattern}. We rewrite these conditions as a pre-processing step. 
Conditions like $upper$ ($S$) = \textit{constant} or $upper$ ($S$) \textit{LIKE pattern},  where the constant or pattern contains at least one lower case character, cannot be satisfied.
Hence for such conditions we do not change the operators but return an empty dataset. 
If these functions are used on a constant string, we convert the string to \textit{upper} or \textit{lower} according to the function. 

\subsection{Solving String Constraints}
\label{sec:string:solver}

There are several available string solvers that we considered, including 
Hampi \cite{HAMPI}, Kaluza \cite{KAL}, SUSHI \cite{SUSHI} and Rex \cite{REX}.
However, we found that Hampi and Kaluza were rather
slow, and while they handled regular expressions and length
constraints, they could not handle constraints such as $S1 < S2$, where both $S1$ and $S2$ are variables.
Rex and SUSHI, though much faster, could not handle constraints involving 
multiple string variables.  Hence, we built our own solver which is described in \supplement{sec:app:string}.
Subsequent to the implementation of our string solver the latest version of CVC (CVC4) has also provided some support for solving string constraints  \cite{cvc4:string}, but it has some limitations currently\footnote{\newstuff{Although there are some limitations in CVC4 currently; in future we may use CVC4 as an integrated solver for both string constraints and other constraints.}}. \newstuff{Refer the experimental section in \supplement{sec:app:string} for details.}

Once the values for string variables are obtained we solve the non-string constraints using CVC3 and
get an overall solution as follows:
enumeration types are created in CVC3 for string variables, with the enumeration names being 
the (suitably encoded) strings generated by the string solver. 
For example, consider a query which has a single string constraint: $S_1$
\textit{like} $`Bio\%'$.  Let the string that satisfies the constraint
be \textit{Biology}, then the constraint is specified as 

\begin{small}
\begin{alltt}
ASSERT(table[index].pos = Biology)
\end{alltt}
\end{small}
in CVC3, where \texttt{table[index].pos} is the corresponding CVC3 variable of $S_1$. 
We then add constraints in CVC3 equating each string variable to its 
corresponding enumeration name, add other non-string constraints as described in Section~\ref{sec:bg} and invoke CVC3 to get a suitable dataset.

If there are disjunctions in the selection predicate, it is not possible to separate the string constraints since not all string constraints may need to be satisfied. 
\eat{
For queries containing disjunctions, in DNF form ($c_1 \vee c_2 \vee ... \vee c_n$, 
where each $c_i$ is a conjunction of predicates),  
each $c_i$ in the constraint is considered one at a time and the string
constraints of that $c_i$ are passed to the string solver. 
If CVC3 can get a solution on incorporating the values returned by the string solver, 
the given solution is sufficient.
If not, we use the next $c_i$. 
If the solver we use is an integrated solver that can handle both string and other constraints 
such issues do not arise since 
we no longer need to separate the the string constraints.

If we are unable to generate a dataset with any $c_i$, the constraints are unsatisfiable.
This method works for single block as well as multi-block queries where there is disjunction in only one query block.
In case there are disjunctions at different levels (e.g. both in outer query block and subqueries) we need to consider combinations of $c_i$ at each level in order to satisfy the constraints. Our current implementation does not support this.}


\subsection{Killing String Constraint Mutations}
\label{sec:string:mutant}

There can be different types of string mutations depending on  whether the string condition is a comparison condition or a LIKE condition.

\paragraph{String Comparison Mutation}~\\ Consider a string constraint of
the form $S1$ \relop\ $S2$, where $S1$ is a variable (attribute name), $S2$ could 
be another variable or a constant. We consider mutations
of \relop\ where any occurrence of one of $\{ =, <>, <, >, \le, \ge\}$ is 
replaced by another. Three datasets are 
enough to kill all the \relop\ mutations. These are the datasets generated for
(1) $S1 = S2$  (2) $S1 > S2$  (3) $S1 < S2$. These datasets will also kill the mutation because of missing string selection mutations. 
In addition, to kill mutations between $=$ and $\sim =$, we generate an additional dataset,
 where $S1 <> S2$, but $S1 \sim = S2$.

\paragraph{LIKE Predicate Mutation}~ \\
We also consider the mutation of the \likeop\ operators where one of 
$\{$\textit{LIKE, ILIKE, NOT LIKE, NOT ILIKE}$\}$ is mutated to another or the operator is missing.
For a condition $S1$ \likeop\ \textit{pattern}, where $S1$ is an attribute name, 
the three datasets given below are sufficient to kill all mutations among the LIKE 
operators: \\
\noindent \textbf{ Dataset 1} satisfying the condition $S1$ \textit{LIKE} \textit{pattern}.\\
\noindent \textbf{ Dataset 2} satisfying condition $S1$ \textit{ILIKE} \textit{pattern},
but not $S1$ \textit{LIKE} \textit{pattern}. \\
\noindent \textbf{Dataset 3} failing both the \textit{LIKE} and \textit{ILIKE} conditions.

\begin{table}
\begin{center}
\begin{tabular}{|l|c|}
\hline
\textbf{Mutation to kill} & \textbf{Dataset}\\ \hline
LIKE vs.\ NOT LIKE & 1, 2, 3\\ \hline
LIKE vs.\ ILIKE & 2\\ \hline
LIKE vs.\ NOT ILIKE & 1, 3\\ \hline
NOT LIKE vs.\ ILIKE & 1, 3\\ \hline
NOT LIKE vs.\ NOT ILIKE & 2\\ \hline
ILIKE vs.\ NOT ILIKE & 1, 2, 3\\ \hline
Missing LIKE / ILIKE & 3 \\ \hline
Missing NOT LIKE / NOT ILIKE &1 \\ \hline
\end{tabular}
\caption{Dataset required to kill like operator mutations}
\label{table:string:mutation}
\end{center}

\end{table}

For example, for the condition  S1 \textit{LIKE} \textquoteleft bio\_\textquoteright, 
the conditions in the three cases would be
(1) S1 \textit{LIKE} \textquoteleft bio\_\textquoteright,
(2) S1 \textit{LIKE} \textquoteleft BIO\_ \textquoteright,
and  (3) S1 \textit{LIKE} \textquoteleft CIO\_\textquoteright.

The targeted mutations and the datasets that kill them are shown in Table~\ref{table:string:mutation}. 


\paragraph{LIKE Pattern Mutations}~\\
A common error while using the LIKE operator is the specification of an
incorrect pattern in the query, for example, specifying $S_1$ LIKE `Comp\_' or $S_1$ LIKE `Com\%' in place of $S_1$ LIKE `Comp\%'. 
There could be a very large number of such patterns to be considered. We handle mutations that involve `\_' in place of `\%' and \textit{vice versa} and also missing `\_' or `\%'. \newstuff{Consider the like predicate to be $S$ \likeop\ P.}
\begin{itemize}
 \item For killing the mutation of `\%' to `\_' or for missing `\%',
we generate separate datasets for each occurrence of the `\%' replaced with ``\_\_''\newstuff{(two underscores)} . 
 The pattern with `\%' gives a non-empty result while the mutated patterns will give an empty result on the corresponding datasets \newstuff{if the \likeop\ is LIKE or ILIKE.
 For NOT LIKE and NOT ILIKE the pattern with `\%' gives an empty result while the mutated patterns will give a non-empty result.}
 \item For killing the mutation of `\_' to `\%' or for missing `\_', 
 we generate separate datasets for each occurrence of `\_' with that occurrence of `\_' removed. 
 \newstuff{If the \likeop\ is LIKE or ILIKE the original pattern gives an empty result while the mutated patterns give non-empty results on the corresponding dataset.
 For NOT LIKE and NOT ILIKE the pattern with `\_' gives a non-empty result while the mutated patterns will give an empty result.}
\end{itemize}

\eat{
\subsection{Completeness}
The completeness of our techniques for data generation for predicates
involving string predicates is dependent on the string solver.
If a separate solver is used for solving string constraints (as our string solver) our data generation techniques are complete provided string conditions may be separated from other constraints.

For killing string selection mutations a mutation of  one \relop\ to another \relop\ will produce a different result on at least one of three datasets. 
In case it is not possible to generate one of the datasets it is easy to see that the mutations are equivalent.
Similarly, as shown in Table~\ref{table:string:mutation} at least one of the dataset will kill mutations of one \likeop\ to another \likeop.
If data generation for one of the datasets fails the mutations that were to be kill by the corresponding dataset are equivalent. 
For example, if dataset 2 cannot be generated then it is not possible to have a dataset that can distinguish 
between LIKE and ILIKE and hence they are equivalent for the query. This also implies NOT LIKE and NOT ILIKE are equivalent. \reminder{change}
For string pattern mutation the number of possible mutations is very large (possibly infinite). 
Hence we not provide completeness guarantees for string pattern mutations. 
We only consider a set of common errors as defined in \ref{sec:string:mutant} and for those cases our techniques will catch a mutation if the mutation is not-equivalent.
}

 \section{Handling NULLs}
\label{sec:nulls}
In our earlier work \cite{xdata:icde11}, we could not handle NULLs. 
\newstuff{ In this section, we discuss how we model NULLs using regular non-NULL values; 
to the best of our knowledge, none of the SMT solvers supports NULL values with SQL NULL value semantics.}

To model NULLs for string attributes, we enumerate a few
more values in the enumerated type and designate them NULLs.
\fullversion{
The
structure of these values is: NULL\_\textless DOMAIN-NAME\textgreater
\_\textless COUNT\textgreater.
}
For example, the domain of course\_id
is modeled in CVC3 as follows:

\begin{small}
\begin{alltt}
DATATYPE course_id = CS190 | CS632 | NULL_course_id_1
| NULL_course_id_2 END;
\end{alltt}
\end{small}

Here, the first two values are regular values from the domain of
course\_id, while the last two values are used as NULLs.
For numeric values, we model NULLs as any integer in a range of negative values 
that are not part of the given domain of that numeric value.




Next, we define a function which identifies which values are
NULL values and which are not. This function is syntactic sugar for dealing
with NULLs cleanly and is defined per domain to identify the NULLs
in that particular domain.  
In addition to specifying which values are NULLs, we
also explicitly need to state that the other values are NOT NULL.
Otherwise, the solver may choose to treat a NON-NULL value as a NULL value.
Following is an example of the function in CVC3:

\begin{small}
\begin{alltt}
ISNULL_COURSE_ID : COURSE_ID -> BOOLEAN;
ASSERT NOT ISNULL_COURSE_ID(CS190);
ASSERT NOT ISNULL_COURSE_ID(CS632);
ASSERT ISNULL_COURSE_ID(NULL_crse_id_1);
ASSERT ISNULL_COURSE_ID(NULL_crse_id_2);
\end{alltt}
\end{small}

We also need to enforce another property of nulls, namely, that
nulls are not comparable. To do so, we choose different
NULL values for different constraint variables that may potentially
be assigned a null value,
thus implicitly enforcing an inequality between them.

The capability to generate NULLs enables us to handle
nullable foreign keys, selection conditions involving IS NULL checks and kill mutations of COUNT to COUNT(*).

\subsection{Nullable Foreign Keys}
If a foreign key attribute $fk$, is nullable 
then the foreign key constraint is encoded in the SMT solver by
forcing values of $fk$ to be either values from the corresponding 
primary key values or NULL values; this allows the SMT solver to
assign NULLS to foreign keys if required. 
Nullable foreign keys allow us to kill more mutants than is possible
if the foreign key attribute as not nullable.
(Our implementation handles multi-attribute foreign and primary keys.)

\subsection{IS NULL / NOT IS NULL Clause}
If the query contains a condition $R.a$ IS NULL, we explicitly assign 
(a different) NULL to attribute $a$ for each tuple $R[i]$ if the query contains only inner joins or only a single relation (provided the attribute is nullable; attributes declared as primary key or as not null cannot be assigned a NULL value). 

However in case the query contains an outer join there may be multiple ways to ensure that an attribute has NULL value.
Let us consider the join condition E1 $\LOJoin$ E2. If the IS NULL condition is on an attribute of E1 we need to ensure that the value of that attribute is NULL.
If the IS NULL condition is on an attribute on E2 we need to ensure that either (a) that attribute is NULL (which may not be possible if E1 is a relation and the attribute is not nullable) or 
(b) for that tuple in E1 there does not exist any matching tuple in E2; this can be done by a minor change in the algorithm to handle NOT EXISTS subqueries as described in Section~\ref{sec:subq:subq} (Algorithm~\ref{algo:notExists}). We omit details for brevity. 

We consider mutation from IS NULL to NOT IS NULL.
The first dataset (the one that generates non-empty results on the original query) kills 
the mutation of IS NULL to NOT IS NULL if the IS NULL condition is present in the form of conjunctions with other conditions.
In the presence of disjunctions, we generate a dataset such that the IS NULL condition is satisfied while the conditions present in disjunction with the IS NULL condition are not satisfied. 
If the query contains an IS NULL then the dataset will give a 
non-empty result whereas the NOT IS NULL mutant will generate an empty result and vice versa. 
We also consider the mutation where the mutant query does not contain the IS NULL condition
In order to kill this mutation we generate a tuple with the IS NULL condition being replaced by NOT IS NULL (with the conditions present in disjunction with the IS NULL not being satisfied).
The original query gives an empty result while the mutant gives a non-empty result. 

If the query contains the condition NOT IS NULL the corresponding mutations can be killed in a similar manner.

\subsection{NULLs and COUNT(*)}
To kill the mutation from COUNT(\textit{attr}) to COUNT(*), where \textit{attr} is a set of attributes, we create a dataset such that all tuples in a group have \textit{attr} as NULL (provided all attributes in \textit{attr} are nullable and none of them is forced to be non-nullable by selection or join conditions).
COUNT(\textit{attr}) gives a count of 0 while COUNT(*) gives a count of equal to the total number of tuples. 

In order to kill mutations of COUNT(*) to COUNT (\textit{attr}), for any set of attributes \textit{attr}, we create a dataset such that all nullable columns (columns that can be assigned NULL values and do not have conditions that force them to be not NULL) have NULL values. If any attribute in \textit{attr} is not nullable, COUNT(*) and COUNT(\textit{attr}) are equivalent mutations. 

 \section{Constrained Aggregation}
\label{sec:constrainedagg}

In \cite{xdata:icde11} we considered aggregates which did not have any constraints on the aggregation result e.g. via a HAVING clause, or in an enclosing SQL query of a subquery with aggregation.
In this section, we discuss techniques for data generation for queries which have constrained aggregation.
We assume that each aggregate is on a single attribute, not on multiple attributes or expressions.
We also assume that aggregation constraints do not involve disjunctions.

Consider the HAVING clause constraint, $SUM(r.a) $ $>$  $20$. 
In case the domain of $r.a$ is restricted to [0,5] it is not possible to generate a single tuple for $r$ such that the aggregation constraint is satisfied.
\newstuff{Most constraint solvers including CVC3 do} not support a relation type where the number of tuples may be left unspecified.
\newstuff{Some solvers like Alloy~\cite{alloy} do support a relation type. However, there are other limitations to using Alloy since it is very slow and supports only the integer datatype.}
We model relations as arrays of tuples with a predefined number of tuples in each relation;
such aggregation constraints cannot be 
translated into SMT solver constraints leaving the number of
tuples  unspecified. 
Hence, before generating SMT solver constraints we must
(a) estimate the number of tuples $n$, required to satisfy the aggregation constraints,  and
(b) in case the input to the aggregate is a join of multiple base relations, translate this number $n$ to appropriate number of tuples for each base relation 
so that the join result contains exactly $n$ tuples.

In Section~\ref{sec:cagg:est} we discuss how to estimate the number of tuples to satisfy an
aggregation constraint. \newstuff{We discuss data generation for constrained aggregation on a single relation in Section~\ref{sec:cagg:single} and for join results in Section~\ref{sec:cagg:join}. }
\eat{Later, in Section~\ref{sec:tupleAssgn} we describe our new method showing how the estimated number 
is translated to an appropriate number of tuples for each base relation, in case the 
input to the aggregate is a join of two or more relations. 
In Section~\ref{sec:cagg:dgen} we describe how to generate data satisfying the constraints. }

\subsection{Estimating Number of Tuples per Group} 
\label{sec:cagg:est}

We now consider how to estimate the number of values (tuples), $n$,  needed to satisfy
aggregation constraints. For each attribute, $A$, on which there are aggregate constraints
we consider the following for estimating $n$.

\begin{enumerate}
 \item \textit{Aggregation Properties}: Constraint variables \textit{sum$_A$, min$_A$, max$_A$, avg$_A$, count$_A$} respectively correspond to the results of aggregation operators SUM, MIN, MAX, AVG and COUNT on attribute $A$. Note that \textit{count$_A$} also indicates the number of tuples at the input to the aggregation.  We add the following conditions
    \begin{itemize}
     \item Since the value of each tuple cannot be less than \textit{min$_A$} and greater than \textit{max$_A$}, it 		follows that \textit{min$_A$}$*$\textit{count$_A$} $\le$ \textit{sum$_A$} $\le$ 		\textit{max$_A$}$*$\textit{count$_A$}.
     
     \item If the domain of $A$ is integer and $A$ is unique,\\ \textit{min$_A$} + (\textit{min$_A$ + 1}) + ... + (\textit{min$_A$ + $count_A$ - 1}) $\le$ $sum_A$ $\le$  ($max_A$ - $count_A$ + 1) + (\textit{$max_A$ - $count_A$ + 2}) + ... + (\textit{max$_A$} - $count_A$ + ($count_A$ - 1)) +(\textit{$max_A$}). \newstuff{We use the simplified form of the above expression for constraint generation.}

     \item $($\textit{$avg_A$}$*$\textit{$count_A$}$) = $ \textit{sum$_A$}

    \end{itemize}

 \item \textit{Domain Constraints}: Constraint variables \textit{dmin$_A$, dmax$_A$} correspond to the minimum and maximum value in the domain of attribute $A$. We add the following constraints
    \begin{itemize}
     \item \textit{dmin$_A$} $\le$ \textit{min$_A$} $\le$ \textit{max$_A$} $\le$ \textit{dmax$_A$}.
	This constraint states that \textit{min$_A$} cannot be less than the domain minimum or 
	greater than \textit{max$_A$}. 
    \end{itemize}

    \item \textit{Aggregation Constraints}: Aggregation constraints specified by the query (e.g. \textit{sum$_A$} $\le$ $10$).
 \item \textit{Selection Conditions}: If the query contains non-aggregate 
 constraints on any attribute $A$, we add these to the tuple estimation constraints. 
For example, consider the query,

\begin{small}
\begin{alltt}
SELECT dept_name,SUM(credits) FROM
course INNER JOIN dept USING (dept_name)
WHERE credits <= 4 GROUP BY dept_name
HAVING SUM(credits) < 13
\end{alltt}
\end{small}

Here, because credits column has a selection
condition on it, its limit is constrained. Hence, \textit{max}$_{credits}$ $\le4$ is
also  added to the list of constraints above.

\end{enumerate}

The solver returns a value for the count which satisfies all the constraints above, but
the value may not be the minimum.  
Since we are interested in small datasets, we want the count to
be as small as possible. Hence, we run CVC3 with the count fixed to different values,
ranging from $1$ to \textit{MAX\_TUPLES} and choose the  smallest value of the count for 
which CVC3 gives a valid answer.\footnote{Since we are interested in 
small datasets, we set \textit{MAX\_TUPLES} to $32$ in our experiments.}  
We borrow the idea of calculating the number of tuples, using multiple tries, for the
aggregation constraint from RQP~\cite{RQP}.   However, note that the problem
is different here, since, unlike RQP, we do not know the value of
the aggregation in the  query result. 
Note that the above procedure works even in case of multiple aggregates on the same column or on different columns.

\paragraph{Heuristic Extensions} ~\\
The value with which the aggregate is compared to may be a column (i.e.\ a variable) e.g.\ HAVING SUM(R.a) \relop\ S.b. 
This can happen when S.b is a group by attribute or when the constrained aggregation is in a subquery and S.b is a correlation variable from an outer query. 
For such cases, we replace the column name by a CVC3 variable when estimating the number of tuples.
We also add the domain and selection conditions for that column as constraints on the CVC3 variable. 
The solver then chooses a value for the number of tuples such that the aggregate is satisfied for some value of the variable in its domain. 

If the aggregation has a DISTINCT clause we add constraints to make the corresponding aggregated attribute unique. 

Handling constraint aggregation in general for these cases is an area of future work.

\subsection{Data Generation for Aggregate on a Single Relation}
\label{sec:cagg:single}
In case the aggregate is on a single relation the number of tuples estimated is assigned to the only relation.
For each result tuple generated by an aggregation operator, 
we create a tuple of constraint variables where group by attributes are 
equated to the corresponding values in the inputs and aggregation 
results are replaced by arithmetic expressions.
For example,
$sum(r.x)$ is replaced by $R[i].x + R[i+1].x + \ldots + R[i+k].x$, 
where R[i] to R[i+k] are the tuples assigned for a particular group. 
We also add constraints to ensure that no two tuples in R[i]...R[i+k] are the same if the relation has a primary key. 

The tuple variables created as above can be used for other operations e.g. selection or join that use the aggregation result as inputs.

In case data generation for multiple groups
is required we add constraints to ensure that at least
one of the GROUP BY attributes is distinct across groups.

Consider the query,

\begin{small}
\begin{alltt}
SELECT id, COUNT(*) FROM takes
WHERE grade = `A+' AND year = 2010
GROUP BY id HAVING COUNT(*) < 3
\end{alltt}
\end{small}
For this query the number of tuples in the group is estimated to be 1. 
We assign a single tuple to the \texttt{takes} relation and add constraints to ensure that \texttt{grade} for this tuple is `A+' and \texttt{year} is 2010.

Note that if the XData system generates additional tuples for the \texttt{takes} relation 
(for example because this query is part of a subquery and there may be other instances of the \texttt{takes} relation outside the subquery or takes is referenced by some other relation and we need to generate additional tuples to satisfy foreign key dependencies) 
the value of COUNT() in the having clause may change and the constrained aggregation may no longer be satisfied.
In order to ensure that the HAVING clause is not affected  we need to ensure that no other tuple in the \texttt{takes} relation belongs to the same group.

In general, to ensure that the additional tuples generated do not cause problems we add 
constraints to ensure that for any additional tuple either has a different value for the GROUP BY attribute and 
hence belongs to a different group or fails at least one of the selection conditions. 
In the above example, we assert that either the \textit{id} is different for the additional tuple or $year \neq 2010$.
\newstuff{In practice, the conditions are generated by Algorithm~\ref{algo:noExtraTuples} described in \supplement{sec:app:noExtra}, which handles the general case of aggregation on join results, to assert these constraints as described in Section~\ref{sec:cagg:dgen}.}

\eat{
In cases where there are no GROUP BY attributes and there are no selection conditions this technique does not work. 
For some special cases like $SUM>val$ (for positive values), $MIN<val$, $MIN<=val$, $MAX>val$ and $MAX>=val$, 
where $val$ is a constant the aggregate constraint will be satisfied irrespective of whether 
there are any additional tuples generated or not. 
For cases like $MIN>val$, $MIN>=val$, $MAX<val$, $MIN<=val$, $AVG>val$, $AVG<val$ 
we add constraints to ensure that any tuple that is added is $>val$, $>=val$,  $<val$, $<=val$, $>val$ and $<val$ 
respectively if it belongs to same group. 
Handling such cases in general is an area of future work. 
}

\subsection{Data Generation for Aggregation on JOIN Results}
\label{sec:cagg:join}
In case the aggregate is on a join result we need to assign tuples to each of the relations such that 
the join results in the required number of tuples.
In this section, we address this issue.

\subsubsection{Estimating Number of Tuples per Relation}
\label{sec:tupleAssgn}
We assume here that all join conditions are equijoins. The required number of tuples is denoted by $n$.
Consider a query that involves $R_i \Join R_j \Join R_k$ where we need $n$ tuples for a GROUP BY on $A.a$. 
Each of the relations need be assigned a specific number of tuples such that the result of the join produces $n$ tuples.

A naive  way is to assign $n$ tuples to a relation, 
$R_i$ and assign the same value to all its joining attributes, $\{R_i.a,R_i.b\}$. 
For relations joining with $R_i$ only a single tuple is assigned and the joining attribute(s) are assigned the value to the corresponding join attribute of $R_i$.
For all other relations also single tuple is assigned and the joining attributes are equated. 
It is easy to see that this assignment will lead to $n$ tuples in the output. 
The assignment, however, does not work in case the joining attribute(s) of $R_i$ are unique (either due to primary keys or by 
inference from other primary keys) or multiple values are required for attributes of some other relations (to satisfy the aggregate constraint). 


We define the following types of attribute(s) that are used  for assigning cardinality to relations.
\begin{enumerate} \itemsep 0em
 \item \textit{uniqueElements}: Sets of attributes for which no two tuples in a group can have the same value. 
These sets of attributes are placed in \textit{uniqueElements}, where \textit{uniqueElements}[$R_{i}$] contains sets of unique elements of relation $R_{i}$. If a relation $R_i$ has unique constraints for (a,b) and (a,c) then \textit{uniqueElements}[$R_{i}$]=\{\{a,b\},\{a,c\}\}.
\item \textit{singleValuedAttributes}: The attribute(s) which have the same value across all tuples in a group. These attributes are placed in \textit{singleValuedAttributes}. 
\end{enumerate}

\newstuff{
Using the \textit{uniqueElements},  \textit{singleValuedAttributes}, join conditions and foreign key conditions for each relation under conditions we estimate the number of tuples for each relation. Details for this are provided in the Appendix~\ref{sec:app:cardinality}.}

\subsubsection{Data Generation}
\label{sec:cagg:dgen}

After getting the tuple assignment for each relation we add CVC3 constraints to fix the number of tuples in a group to the estimated value. 
For each join condition, constraints are generated depending on 
the number of tuples assigned. For example, if both relations $R$ and $S$ have $n$ tuples,
the constraint $R[i].x = S[i].y$ is generated for all $1 \leq i \leq n$,
while if $R$ has $n$ tuples and $S$ has 1  tuple, the constraint
$R[i].x = S[1].y$ is generated for all $1 \leq i \leq n$.
Constraints variables for the output of the aggregate operator are created as described earlier in Section~\ref{sec:cagg:single}. One difference is in handling aggregation for relations that have been assigned one tuple.  
For example,
$sum(r.x)$ is replaced by $R[i].x + R[i+1].x + \ldots + R[i+k].x$, where R[i] to R[i+k] are the tuples assigned for a particular group,
if $R$ has $n$ tuples, otherwise it is replaced by $n*R[i].x$, where $R[i]$ is the only tuple assigned for a group. 
Unique constraints are added as pairwise non-equality constraints to ensure that sets of $uniqueElements$ have distinct values.

\newstuff{
Constraints to ensure that additional tuples do not alter satisfaction of the aggregate conditions for the group $g$ are generated using 
Algorithm~\ref{algo:noExtraTuples} described in \supplement{sec:app:noExtra}. 
The inputs to the algorithm are (a) $T$ - query tree corresponding to block that contains the constrained aggregate (b) $AT$ - the tuples generated to satisfy the  constrained aggregation and (c) $ASel$ - conditions that evaluate each GROUP BY attributes to the corresponding values in $g$.
}

Data generation for multiple groups is done by  adding constraints to ensure that at least
one of the GROUP BY attributes is distinct across groups.

The constraints are then given as input
to CVC3, and output of CVC3 gives us the required dataset.

\paragraph{Discussion:}~\\
Our tuple assignment techniques always assign either 1 tuple or $n$ tuples to a relation.
There could be cases where such an assignment is not possible and a different assignment is required to generate datasets.
However, in such an assignment it becomes difficult to assert constraints 
such that the join of the relations will generate exactly the required number of tuples.
Handling tuple assignment for cases where either 1 or $n$ tuples cannot 
be assigned to all the relations to satisfy the aggregation constraint is an area of future work.

\subsection{Constraint Aggregation and Mutant Killing}
 
\newstuff{Techniques for killing aggregation mutations were described in 
\cite{xdata:icde11} (summarized in Section~\ref{sec:bg:mutation}). 
A dataset to kill aggregation mutations is generated by creating multiple tuples per group using techniques of 
constrained aggregation described above. 
Different mutations of the aggregate operator will produce different values  on this dataset.
To ensure that the value difference due to aggregate mutation will cause a difference in the constraint aggregate result,
we need to ensure that only one of the query or its mutation satisfies the aggregation constraint.
For some cases, we have implemented constraints to ensure that there is a difference in the constraint aggregate result.
Implementing this in general is an area of future work.
}

\newstuff{Datasets for killing mutations of comparison operators in aggregation constraint (e.g. having clause) are generated using existing techniques 
in XData for handling comparison mutations. Killing mutations due to additional and missing group by attributes is discussed in Section~\ref{sec:group}. }

 \section{Where Clause Subqueries} 
\label{sec:subquery}
We now consider test data generation for SQL queries involving subqueries. 
Data generation for subqueries in the FROM clause is discussed in Section~\ref{sec:misc}; in this section we consider data generation and mutation killing for subqueries in the where clause.
We initially assume in Section~\ref{sec:subq:subq} that subqueries do not have aggregations. 
Subqueries with aggregation  are discussed in Section~\ref{sec:subq:agg}.

\subsection{Data Generation for Subqueries Without Aggregation}
\label{sec:subq:subq}
\paragraph{EXISTS Connective}~\\
Consider a query Q with a nested subquery predicate \texttt{EXISTS($SQ$)}.
To generate a non-empty result for Q we need to ensure that SQ gives a non-empty result.
If SQ does not have any correlation variables we treat subquery SQ as a query in itself and add constraints to generate a non-empty dataset for the subquery using our data generation techniques. We then add constraints for Q for predicates other than the subquery. The dataset is then generated based on these constraints.

If SQ has correlation conditions, then for every tuple that is generated for Q, we call a function to generate the constraints for data generation of the subquery, with the correlation variables passed as parameters. 
The correlation conditions are treated as selections in SQ
with the given constraint variables and appropriate constraints are generated for SQ.
For example, consider the query \\
\begin{small}
\verb|SELECT course_id,title|\\
\verb|FROM course INNER JOIN section USING(course_id)|\\
\verb|WHERE year = 2010 AND EXISTS (SELECT * FROM prereq|\\
\verb|WHERE prereq_id=`CS-201' AND|\\
\verb|prereq.course_id = course.course_id)|\\
\end{small}

To generate a dataset for the outer query, we generate a single tuple each for the course and section relations. Let the tuples be \textit{course[1]} and \textit{section[1]}. We then add constraints to assert 
\textit{section[1].year=2010} and 
\textit{course[1].c\-ourse\_id = section[1].course\_id}. 
We pass the correlation variable \textit{course[1].course\_id} as a parameter to the function for generating constraints for the subquery. 
For this tuple in the outer query block, we generate a tuple in prereq relation, say \textit{prereq[1]}, 
for which we add constraints to ensure that \textit{prereq[1].prereq\_id = `CS-201'} 
and \textit{prereq[1].course\_id = course[1].course\_id}.

\paragraph{NOT EXISTS Connective}~\\
Consider a query Q with a nested subquery predicate \texttt{NOT EXISTS($SQ$)}.
Here we need to ensure that the number of tuples from SQ is 0. 

If SQ has only a single relation R, we add constraints to ensure that every tuple in R fails at least one of the selection conditions.
In case, SQ has a join of two or more relations we traverse the tree of SQ, and in a recursive manner 
add constraints on selections and joins to ensure that no tuple reaches the root of SQ.
If the join is an INNER JOIN we need to ensure that there  exists no pair of tuples for which 
the join conditions are satisfied or that one of the inputs to the join is empty. 
In case the join is LEFT OUTER JOIN, we need to ensure that there is no tuple in the left subtree. 
Similarly, in case of RIGHT OUTER JOIN we need to ensure that no tuple is projected from the right subquery.

\begin{algorithm}
  \renewcommand{\algorithmicrequire}{\textbf{Inputs:}}
    \renewcommand{\algorithmicensure}{\textbf{Output:}}
\begin{algorithmic}[1]
\REQUIRE $T$ = Query tree
\ENSURE constraints to ensure no tuple is projected from the subquery 

\STATE constraints $\leftarrow$ ``~''
\STATE $R$ = $T$.ROOT
\IF{$R$ is a relation}
  \STATE Let the selection conditions on $R$ be $S_1$ $\wedge$ $S_2$ $\wedge$..$S_c$
  \STATE Let the number of tuples in $R$ be $m$
  \FOR{i in 1 to m}
    \STATE SC[i] $\leftarrow$ NOT($S_1$) OR NOT($S_2$) .. OR NOT($S_c$)
  \ENDFOR
  \STATE constraints $\leftarrow$ SC[1] AND .. SC[i] AND SC[m] \\$\forall$ $i, 1 \leq i \leq m$
\ELSIF{$R$ is an aggregate}
  \STATE genConstraintsForNotExists($R$.CHILD)
\ELSIF{$R$ is a LEFT OUTER JOIN}
  \STATE constraints $\leftarrow$ genConstraintsForNotExists($R$.LEFT)
\ELSIF{$R$ is a RIGHT OUTER JOIN}
  \STATE constraints $\leftarrow$ genConstraintsForNotExists($R$.RIGHT)
\ELSIF{$R$ is an INNER JOIN}
  \STATE JC=\{\}
  \STATE Let the join conditions at R be $J_1,J_2..J_c$
  \FOR{Each join condition $J_k$}
    \STATE Let $J_k$ involve relations $R_1$ and $R_2$
    \STATE Let the number of tuples in $R_1$ be $m$ and in $R_2$ be $n$
    \STATE Let $J_k(i,j)$ denote the condition corresponding to join of tuples $R_1[i]$ and $R_2[j]$
    \FOR{i  in 1 to m, j in 1 to n}
	\STATE JC[k]~$\leftarrow$~JC[k] AND NOT($J_k(i,j)$)
  \ENDFOR
  \ENDFOR
  \STATE constraints$\leftarrow$ JC[1] OR..  JC[k] OR.. JC[c]\\ $\forall k, $ $1 \leq k \leq c$
  \STATE constraints $\leftarrow$ constraints + ``OR'' + (genConstraintsForNotExists($R$.LEFT)) 
  + ``OR'' + (genConstraintsForNotExists($R$.RIGHT))
\ENDIF

\RETURN{constraints}
\end{algorithmic}
\caption{ : genConstraintsForNotExists}
\label{algo:notExists}
\end{algorithm}

\newstuff{Constraints to ensure that there is no tuple from the NOT EXISTS subquery are added 
using Algorithm~\ref{algo:notExists}.}
\newstuff{If the subquery contains selections with disjunctions, we may fail to get the selection conditions that involve only $R$ in Step~4 of our algorithm.
Our algorithm is currently restricted to NOT EXISTS queries that do not contain any disjunction.
At Step~5 we assert negations of the constraints corresponding to the particular selection condition, $S_i$. For example, if $S_i$ is  a NOT EXISTS subquery we assert constraints corresponding to EXISTS($S_i$).}
Correlation variables in SQ, if present, are treated in the same manner as EXISTS subquery and passed as parameters. Correlation conditions are then treated as selections in Algorithm~\ref{algo:notExists}.

\paragraph{IN/NOT IN Connective}~\\
We convert subqueries of the IN type to EXISTS type subquery by adding the IN connective as a correlation condition in the WHERE clause of the EXISTS subquery. The same techniques as that of EXISTS are then used. 
Similarly. subqueries using a NOT IN connective are converted to use the NOT EXISTS connective.
For example, \\
\begin{small}
\verb|r.a IN (SELECT s.b FROM .. WHERE ..)|
\end{small}
\\is converted to\\
\begin{small}
\verb|EXISTS (SELECT s.b FROM .. WHERE .. AND r.a = s.b)|
\end{small}

\paragraph{ALL/ANY Connective}~\\
Subqueries with ALL and ANY connectives, always appear with one of the 
comparison operators, for example ``$<$ ALL'' or ``$>=$ ANY''.
We transform subqueries  of the form $relop$ ANY to an EXISTS query with $relop$ condition as a correlation condition in the WHERE clause. 
Subqueries with $relop$ ALL are transformed to a NOT EXISTS query with a negation of the $relop$ condition as a correlation condition, \newstuff{or either of the correlation variables in the correlation condition as NULL} in the WHERE clause.  
For example,\\\begin{small} 
\verb|r.a >ALL (SELECT s.b FROM .. WHERE ..)|\end{small}
\\is converted to \\ \begin{small}
\verb|NOT EXISTS| \\
\verb|(SELECT s.b FROM .. WHERE .. AND r.a <= s.b| \\
\verb|   OR IS NULL(r.a) OR IS NULL(s.b))| \end{small}

\paragraph{Scalar Subqueries}~\\
Scalar subqueries are subqueries that return only a single result. We consider scalar subqueries in the where clause which are used in conditions on the form \texttt{SSQ \relop\ attr/value}, where \texttt{SSQ} is a scalar subquery, \texttt{attr} is an attribute from the outer block of query and \texttt{value} is a constant. 
For scalar subqueries, we generate only a single tuple for the query
and assert that the projected attribute satisfies the comparison operator. 
Correlation conditions, if any, are treated in the same manner as subqueries with the EXISTS connective.

\subsection{Data Generation for Subqueries With Aggregation}
\label{sec:subq:agg}
In this section, we consider subqueries that have aggregation. Constraints involving aggregation can be in the inner query (e.g. HAVING clause) or in outer query (e.g. r.s $<$ (SELECT agg(s.b...)))
\paragraph{Non Scalar Subqueries}~\\
The techniques in Section~\ref{sec:subq:subq} can be applied for EXISTS subqueries without constrained aggregation, since we only need to ensure empty / non-empty results for the subquery. 
For NOT EXISTS Algorithm~\ref{algo:notExists} covers the case of aggregate operators as well.

In case of constrained aggregation in EXISTS subquery (e.g. HAVING clause), we use the techniques described in Section~\ref{sec:constrainedagg} to generate tuples for the subquery; multiple tuples may be generated. 
In case there is a constrained aggregation in the NOT EXISTS subquery, we assert  constraints to ensure that either the constraint aggregation is not satisfied or there is no tuple input to the aggregation constraint.


Subqueries of the IN/NOT IN/ALL/ANY type having an aggregate as the projected attribute can be transformed into EXISTS/NOT EXISTS in a similar manner as shown in Section~\ref{sec:subq:subq}. In this case, the projected aggregate is added as a HAVING clause.
For example,

\begin{small}
\begin{alltt}
r.a NOT IN (SELECT agg(s.b) FROM .. WHERE .. )
\end{alltt}
\end{small}
is converted to 

\begin{small}
\begin{alltt}
NOT EXISTS (SELECT agg(s.b) FROM ..
WHERE .. HAVING agg(s.b) = r.a)
\end{alltt}
\end{small}
The techniques for constrained aggregation in EXISTS/ NOT EXISTS can then be applied.

\paragraph{Scalar Subqueries}~\\
Consider the following query involving the relation \textit{takes}(\textit{id, course\_id, sec\_id, semester, year, grades}),

\begin{small}
\begin{verbatim}
SELECT id FROM takes 
WHERE grade < (SELECT MIN(grade) 
FROM takes WHERE year = 2010)  
\end{verbatim}
\end{small}

To generate datasets for this query we add constraints to generate a tuple, $takes[1]$ for the \texttt{takes} relation in the outer query. 
The tuple estimation technique for the subquery estimates that one tuple is required 
to satisfy the comparison operator ($< MIN(grade)$).
We add constraints to generate one more tuple, say $takes[2]$ for takes relation corresponding to the subquery and add a constraint to ensure that \textit{takes[2].year = 2010} for that tuple.
We then add the constraint, \textit{takes[1].grade} $<$ \textit{takes[2].grade} to ensure that the grade of the outer query tuple is greater than the grade of the subquery tuple.
Since takes[1] does not participate in aggregation we need to ensure that it does not satisfy the conditions of the subquery block. To ensure this,  the constraint $takes[1].year <> 2010$ is added.

In general, consider a query of the form 

\begin{small}
 \begin{alltt}
SELECT * FROM rel1 JOIN .. WHERE cond1 AND ... 
AND attr1 \relop\ (SELECT agg(sqrel1.attr2)
FROM sqrel1 JOIN ... WHERE sqcond1 AND ..) 
\end{alltt}
\end{small}

For such subqueries 
we need to
ensure that the aggregate, \texttt{agg(sqrel1.attr2)} satisfies the condition \texttt{attr1 \relop\ agg(sqrel1.attr2)}. 
In order to do this, we may need to project multiple tuples from the subquery. 
We use the techniques described in Section~\ref{sec:constrainedagg} to estimate the number of tuples, assign the desired number of tuples to each relation and generate constraints for data generation.
\newstuff{In order to ensure that no additional tuple affects the aggregate value, we use the techniques described in 
Algorithm~\ref{algo:noExtraTuples} in \supplement{sec:app:noExtra}. The input to the algorithm is the same 
as described in Section~\ref{sec:subq:mutation} below.}

Similar to EXISTS subquery, in the presence of correlation conditions, we generate one group of tuples in the subquery for every tuple in the outer query.

\subsection{Killing Subquery Connective Mutations}
\paragraph {EXISTS/NOT EXISTS, IN/NOT IN Mutation}~\\
The dataset generated for the original query will kill the mutation between IN and NOT IN, and between EXISTS and NOT EXISTS if the subquery condition is present in the form of conjunctions with other conditions.
In the presence of disjunctions, we generate a dataset such that the subquery condition is satisfied and conditions in disjunction with it are not.
The EXISTS clause gives an empty result when NOT EXISTS gives a non-empty
result, and vice versa. Similar datasets are generated to kill mutations of IN vs.\ NOT IN.

\paragraph{Comparison Operator Mutation}~\\
For conditions of the form ``r.A \relop\ (SSQ)''
where SSQ is a scalar subquery, as well as conditions of 
``r.A \relop\ [ALL/ANY] SQ'',
we consider mutations among the different $\relop$s.
Similar to the approach shown in Section~\ref{sec:bg:mutation} we generate data for three cases,
with \relop\ replaced by $>$, $=$ and $<$. 

\paragraph{ANY/ALL Mutation}~\\
This mutation involves changing from ANY to ALL or vice versa.
Since the ANY subquery has been transformed to EXISTS
the mutation from ANY to ALL becomes a double mutation - replacing EXISTS with NOT EXISTS and negating the correlation condition corresponding to the ANY comparison condition.
The case for ALL to ANY mutation is similar.

Let the correlation condition added because of transformation of ALL/ANY to EXISTS/NOT EXISTS be  $R_1.a$ \relop\ $R_2.b$.  
We generate a dataset with two tuples in the subquery for every tuple in the outer query. 
We add constraints for \relop\ for one tuple and the negation of \relop\ for the other tuple.
The ANY query will produce a non-empty result 
while the ALL query will produce an empty result.

\paragraph{Missing Subquery Mutation}~\\
To kill the mutation of a query with missing EXISTS condition connective we generate a dataset with the EXISTS condition replaced by NOT EXISTS. 
If the EXISTS condition is missing the mutant query will give a non-empty answer while the original query will give an empty answer. 
Similarly, for killing mutations with missing subquery connectives in other cases we replace NOT EXISTS with EXISTS, IN with NOT IN and NOT IN with IN.

The datasets generated to kill comparison operator mutation will also kill mutations involving missing scalar/ALL/ANY subqueries. 
If the subquery is present the original query will give an empty result on at least one of the three datasets while the mutant query will produce a non-empty result on all the three datasets.

\subsection{Killing Mutations in a Subquery}
\label{sec:subq:mutation}
We also need to generate test data for killing mutations in subquery blocks. 
For the EXISTS connective and for scalar subqueries we treat a subquery block as a normal query and 
generate sets of constraints to kill mutations in the subquery block. 
For each constraint set, we also add constraints to ensure
a non-empty result on the outer query block. 

For killing selection (comparison mutations, string mutations, NULL mutations), JOIN, and HAVING 
clause mutations the techniques described in \cite{xdata:icde11} and this paper generate datasets 
that produce empty result on either the query or the mutant but not both. 
Thus for these mutations the subquery will satisfy the EXISTS condition or the comparison operator 
(for scalar subqueries) for either the subquery or its mutation enabling XData to kill the mutation. 

\newstuff{Extra tuples may get generated for the subquery if there are relations in the subquery that 
are repeated in the query or are referenced by other relation through foreign keys. 
Because of these extra tuples, an empty result may turn into a non-empty result or vice versa.
To prevent this, we add constraints using  Algorithm~\ref{algo:noExtraTuples} 
described in \supplement{sec:app:noExtra} where (a)$T$ - query tree of the subquery 
(b)$AT$ - tuples created for the subquery (c)$ASel$ - correlation conditions with 
correlation variables being passed as parameters. 
The constraints ensure that the extra tuples do not affect the result of the subquery, preventing the extra tuples from turning an empty result into a non-empty result or vice versa.}

In case there are disjunctions with the subquery, we add constraints to ensure that other 
conditions in disjunction with the subquery (e.g. \texttt{P or EXISTS(Q)}) are not 
satisfied as described in Section~\ref{sec:disjunct}.

Mutations like distinct or aggregation mutation in the project clause of the subquery 
create  equivalent mutants of the query and hence need not be killed.

If the subquery uses the NOT EXISTS connective, we generate the datasets for killing mutations 
in the subquery treating the NOT EXISTS  as an EXISTS conditions. 
Out of the original query and the mutant, the query that produces empty results on the subquery satisfies the NOT EXISTS conditions and produces non-empty results for the outer query. The query that does not produce empty results does not satisfy the NOT EXISTS condition and produces an empty result in the outer query. Thus, these mutations can be killed. 

Subquery connectives IN, NOT IN, ANY and ALL are converted to EXISTS and NOT EXISTS as described earlier. Mutations in the subquery are killed after the conversion.

\begin{table*}
\begin{center}

\begin{tabular}{|c||c|c||c|c|c|c|c|c|}

\hline  \textbf{Dataset} & \textbf{P} & \textbf{Q} & \textbf{UNION} & \textbf{UNION}
		& \textbf{INTERSECT} & \textbf{INTERSECT} & \textbf{EXCEPT} & \textbf{EXCEPT}\\

	& & & & \textbf{ALL} & & \textbf{ALL} & & \textbf{ALL}\\

\hline 1 & $t_1$& $t_1$  & $t_1$ & $t_1,t_1$ & $t_1$ & 
$t_1$& $\nexists t_1$ &  $\nexists t_1$ \\ 

\hline 2 & $t_1$ & $\nexists t_1$ & $t_1$& $t_1$ & $\nexists 
t_1$ & $\nexists t_1$& $t_1$& $t_1$\\

\hline 3 & $ t_1$  & $\exists ^{>1} t_1$ & $t_1$ & $\exists ^{>1} t_1$ & 
$t_1$ &$t_1$ &$\nexists t_1$ & $\nexists t_1$ \\ 

\hline 4 & $\nexists t_1$ & $t_1$ & $t_1$ &  $t_1$ & $\nexists 
t_1$& $\nexists t_1$&$\nexists t_1$ &$\nexists t_1$\\

\hline 5 & $\nexists t_1$ & $\exists ^{>1} t_1$  & $t_1$ & $\exists ^{>1} t_1$ & 
$\nexists t_1$ &$\nexists t_1$ &$\nexists t_1$ &$\nexists t_1$ \\

\hline 6 &$\exists ^{>1} t_1$ & $t_1$  & $t_1$ & $\exists ^{>1} t_1$ (sum)& 
$t_1$& $t_1$&$\nexists t_1$ & $\exists t_1$ (diff)\\ 

\hline 7 &$\exists ^{>1} t_1$ & $\nexists t_1$  & $t_1$ & $\exists ^{>1} t_1$ & 
$\nexists t_1$ &$\nexists t_1$ &$t_1$ &$\exists ^{>1} t_1$ \\

\hline 8 & $\exists ^{>1} t_1$ & $\exists ^{>1} t_1$  & $t_1$ & $\exists ^{>1} 
t_1$ (sum) & $t_1$& $\exists ^{>1} t_1$ (min) &$\nexists t_1$  & - (diff) \\

\hline

\end{tabular}
\caption{Datasets to kill set operator mutations}
\label{tab:set:datasets}
\end{center}
\end{table*}

\section{Set Operators}
\label{sec:set}
In this section, we consider data generation and mutation killing for queries that contain one of the
following set operators - UNION, UNION ALL, INTERSECT, INTERSECT ALL, EXCEPT, EXCEPT ALL.

\subsection{Data generation}
\label{sec:set:datagen}
Set queries are of the form,
\begin{small}\verb|P SETOP Q|
\end{small}
where \texttt{SETOP} is a set operator, and P and Q are queries that may be simple or compound queries themselves. 

In order to generate a dataset that produces a non-empty result on this query if the \texttt{SETOP} is UNION(ALL) we add constraints to ensure non-empty results for P or Q or both (P and Q may have conflicting constraints so for both to have non-empty results may not always be possible).

Data generation for INTERSECT(ALL) is done in a similar manner as the EXISTS subquery described in Section~\ref{sec:subq:subq}. We treat the query as 

\begin{small}
\begin{alltt}
SELECT * FROM (P) WHERE EXISTS 
    (SELECT * FROM Q WHERE pred)  
\end{alltt}
\end{small}
where predicate \texttt{pred} equates each projected attribute of P to the corresponding attribute of Q. 
For each tuple in P, we generate a corresponding tuple in Q that satisfies the correlation condition, \texttt{pred}, as described in Section~\ref{sec:subq:subq}.
Data generation for EXCEPT(ALL) is done in a similar manner using NOT EXISTS instead of 
EXISTS, using the techniques described earlier for the NOT EXISTS operator.

\subsection{Killing Set Operator Mutations}
\label{sec:set:set_mutation}

In order to kill the mutations among the different operators (UNION(ALL), INTERSECT(ALL), 
EXCEPT (ALL)) we generate  datasets as described below (summarized in 
Table~\ref{tab:set:datasets} along with the results for various set operators). 
\begin{enumerate}
 \item Generate a dataset that has exactly one tuple $t_1$ for P.
 Add constraints to ensure that one matching tuple exists in Q.

 \item Generate a dataset that has one tuple $t_1$ for P. 
 Add constraints to ensure that $t_1$ does not exist in Q. 

\item Generate a dataset which has at least two identical tuples $\exists ^{>1} t_1$ 
 for Q. Add constraints to create one matching tuple $t_1$ for P.
 
 \item Generate a dataset that has one tuple $t_1$ for Q. 
 Add constraints to ensure that $t_1$ does not exist in P. 

\item Generate a dataset which has at least two identical tuples $\exists ^{>1} t_1$
 for Q. Add constraints to ensure not matching tuples for P.

 \item Generate a dataset that has at least two identical tuples, $\exists ^{>1} t_1$
 for P. Add constraints to ensure that there is exactly one matching tuple $t_1$ in Q.
 
 \item Generate a dataset that has at least two identical tuples, $\exists ^{>1} t_1$
 for P. Add constraints to ensure that $t_1$ does not exist in Q. 

\item Generate a dataset that has at least two identical tuples, $\exists ^{>1} t_1$ 
for both P and Q. 
 
\end{enumerate}

We call kill a mutation between a pair of set operators if for a dataset the results of 
the query as shown in Table~\ref{tab:set:datasets} differ.
Note that it may not be possible to generate some datasets because of query/integrity 
constraints; in particular primary key constraints may prevent generation of datasets 
with duplicates. It may not be necessary to generate all datasets to kill all mutations. 
As an optimization we can stop generation of datasets if we have successfully generated 
at least one of the datasets for killing each of the mutations. 

For both P and Q we have three options; either generate no tuple, one tuple or 
more than one tuple. Table~\ref{tab:set:datasets} exhaustively covers all 
combinations (except for the case where both P and Q are empty, since if both P an Q are 
empty all operators would give an empty result and no mutation would be killed).
Hence, these datasets are sufficient to kill all pairs mutations.
For example the mutation between \smalltt{INTERSECT} and \smalltt{INTERSECT ALL} can 
only be killed if there is more than one matching tuple between P and Q. Dataset 8 
covers this case. 
The only mutation that may be missed is the mutation between \smalltt{EXCEPT ALL} and 
other operators except \smalltt{UNION ALL} since for dataset 8, we cannot guarantee 
whether the result would be $\exists t_1$, $\nexists t_1$ or $\exists ^{>1} t_1$. Dataset 
8 would still be able to kill the mutation between \smalltt{UNION ALL} and 
\smalltt{EXCEPT ALL} since \smalltt{UNION ALL} would produce more tuples in the result 
than \smalltt{EXCEPT ALL}. Hence, if it is possible to only generate dataset 8, mutations 
of other operators with \smalltt{EXCEPT ALL} may not get killed.
\eat{The mutation between \smalltt{EXCEPT} and \smalltt{EXCEPT ALL} 
can be killed by a dataset where P contains more number of matching tuples than Q. 
However, our current implementation is restricted to dataset 6, i.e. generate one tuple 
for Q and generate more than one matching tuple for P.}
In order to provide completeness guarantees for killing mutations involving 
\smalltt{EXCEPT ALL}, we need to generate specific number of 
tuples for P and Q. This is an area of future work.

To ensure that a tuples of one relation does not exist in the other, constraints 
are added using the NOT EXISTS technique described in Algorithm~\ref{algo:notExists} of 
Section~\ref{sec:subq:subq}. To ensure that a tuple in one relation exists in the other, 
we use the EXISTS technique described in Section~\ref{sec:subq:subq}.

To create at least two identical tuples in the result of a subquery, we assert 
constraints to imply that the number of tuples is more than one. Then using the 
techniques described in Section~\ref{sec:constrainedagg} for constrained aggregation we 
estimate the required number of tuples for each base relation. 
We treat the projected attributes in the select clause as the group by attributes in 
constrained aggregation, which ensures that these have the same value across tuples.
Data generation is done using techniques for constrained aggregation described in 
Section~\ref{sec:cagg:single} and Section~\ref{sec:cagg:dgen}.

\subsection{Killing Mutations in Input to Set Operators}
\label{sec:set:mutation}
We also need to kill mutations in the input to the set operator. 
For this, we need to ensure that the effect of the mutation makes a difference in the result of the set operator.

If the set operator is  UNION/UNION ALL  and the mutation to the query is in P, we add constraints to ensure that the mutation in P is killed. 
In addition to ensure that there are no tuples from Q that mask the changes in the result we add constraints similar to NOT EXISTS subquery for Q. Similarly data generation can be done for killing mutations in Q. 

We treat INTERSECT and EXCEPT queries as EXISTS and NOT EXISTS respectively as described earlier. 
Mutations of P can be killed by datasets to kill mutations of the outer query block while the mutations in Q can be killing by killing mutations in the subquery block as described in Section~\ref{sec:subq:mutation}.
 
 \section{Handling Join Condition, Group By Attribute and Distinct Clause Mutations}
\label{sec:join:group:distinct}
\newstuff{
In this section, we describe our techniques to kill missing or additional joins conditions, group by attributes and DISTINCT keyword.
Although our previous work handled joins, group by and distinct clause, these  mutations were not considered.}

\subsection{Missing or Extra Joins Conditions}
\label{sec:joincond}

Consider the tables \texttt{student} (\textit{id, name, dept\_name}), \texttt{course} (\textit{course\_id}, \textit{course\_name} and \textit{dept\_name}) and  \texttt{takes} (\textit{id, course\_id, sec\_id, semester, year}) from the University schema in \cite{dbconcepts2010}. 
Consider the query,

\begin{small}
\begin{alltt}
SELECT course_id,course_name
FROM student INNER JOIN takes ON(id)
INNER JOIN course ON(course_id)
WHERE student.id = 1234
\end{alltt}
\end{small}

\noindent One of the mutations of the query could be because of an additional join condition leading to a mutant query like 

\begin{small}
\begin{alltt}
SELECT course_id,course_name
FROM student INNER JOIN takes ON(id)
INNER JOIN course ON(course_id, dept_name)
WHERE student.id = 1234
\end{alltt}
\end{small}

Such errors are common when using natural joins. For example, if natural join was used in place of {\small\texttt{.. INNER JOIN course ON(course\_id)}} resulting in student.de\-pt\_name being equal to course.dept\_name.

In order to kill such mutations, we do the following. 
Let the relations being joined be $R_{i}$ and $R_{j}$.  
For every attribute $p \in R_{i}$ such that (a) there is an attribute $q \in R_{j}$ with identical names and (b) there is no join condition involving $p$ and $q$ in the original query,  we assert that the values held by the two attributes are not equal. 
The original query without the join condition would give a non-empty result while the mutation would give an empty result. 

\eat{
For the above example, we generate a dataset such as the following:

\noindent student : 
\begin{tabular}{c|c|c} 
id & name & dept\_name \\ \hline
1234 & Alice & EE 
\end{tabular} 

\noindent course :
\begin{tabular}{c|c|c} 
course\_id & name & dept\_name \\ \hline
CS-317 & Database Systems & Comp. Sc. \\ 
\end{tabular} 

\noindent takes :
\begin{tabular}{c|c|c|c|c} 
id & course\_id & sec\_id & semester & year \\ \hline
1234 & CS-317 & 1 & Fall &2014 \\ 
\end{tabular}

\noindent where attribute \textit{dept\_name} which occurs in student and course,
but which are not equated in the query, is assigned different values
in the two tuples.
The mutated query 
would give an empty result for this dataset while the original query gives the result (CS-317, Database Systems).
} \reminder{example removed}

Similarly, there could be mutants such that the mutant query contains some missing join conditions. 
Such mutations can be killed by the datasets that kill join type mutations (\textit{INNER / LEFT OUTER / RIGHT OUTER}) described in Section~\ref{sec:bg:mutation}. 

 \subsection{Group By Clause Mutations}
\label{sec:group}
In this section, we discuss the mutation of the query due to the presence of additional attributes or absence of some attributes in the group by clause.

\subsubsection{Additional Group By Attributes}
Consider the following query, $Q$, to find the number of students taking each course every time it is offered.

\begin{small}
\begin{alltt}
SELECT count(id), course\_id, semester, year 
FROM takes GROUP BY course\_id, semester, year
\end{alltt}
\end{small}
Additional attributes  included in the group by clause such as \textit{section} as shown in the query, $Q_s$, below, could result in an erroneous query. 

\begin{small}
\begin{alltt}
SELECT count(id), course\_id, semester, year 
FROM takes GROUP BY course\_id, semester, year, section
\end{alltt}
\end{small}

To catch such mutations, we generate a dataset for each possible additional group by attribute, with more than one tuple in the group, such that the additional attribute (\textit{section} in this case) has different values for different tuples in the group. 
This ensures that the incorrect query produces multiple groups while the correct one produces only a single group, thereby killing the mutation. 
Note that because of some selection conditions resulting in attributes being single-valued, functional dependencies on group by attributes and equality conditions on group by attributes some of the mutations with additional GROUP BY may be equivalent to the original query.
We do not consider such attributes.

There are situations where the above approach would not work e.g. if the group by is in an EXISTS subquery. \newstuff{The EXISTS condition is satisfied regardless of one or two groups being present.} In such a case if  there is no constrained aggregation the mutation would be equivalent but if there are aggregation constraints the mutation may not be equivalent and needs to be killed. 

If the group has an aggregation  that is constrained, e.g., $SUM(a) > 20$ or $SUM(b) \leq  30$ the number of tuples is assigned based on the aggregation constraint. 
We  try to ensure that the data generated is such that the aggregation constraints of one of the queries, i.e., either of the original
query or of its mutant are satisfied, resulting in a non-empty result on either the original query or its mutation but not both, hence killing the mutation.

Let the group by attributes be $G$.
For each possible additional group by attribute, $g_i$,  we generate up to 2 corresponding datasets. 
In our first attempt, we try to generate two separate groups,  which agree on $G$ but differ in $g_i$, 
such that each group (when grouped by $G,g_i$) satisfies the aggregation constraints, 
but the group containing the union of these tuples (i.e., group by $G$) does not. 
Note that this may not be possible in case the aggregate is of the form  \texttt{SUM(x) $>number$} for values in the positive range or \texttt{COUNT(x) $> number$} etc.
Hence, we also try to generate a dataset such that the combined group satisfies the aggregate but the individual groups do not. If either succeeds, the mutation will be killed.

\subsubsection{Missing Group By Attributes}
Another common error is to miss specifying some of the group by attributes. 
For example, if one misses specifying the attribute, \textit{semester} in the GROUP BY clause but query $Q$ then the resultant query is clearly erroneous.
Such erroneous queries can be easily detected if
the number of attributes projected out is different.

However, that may not be the case for all queries where a group by attribute 
has been missed. For instance, in the above example, if \textit{semester} was not
in the projection list, the missing group by mutation would be harder to catch.
Although rare, we have found such cases when the group by is in a subquery whose result is an aggregation tuple.

We generate  datasets to kill such mutations as follows:
Let $g_{1},g_{2},...g_{n}$ be the group by attributes. 
For missing group by  attribute, $g_i$, we treat the original query as the one with the missing group by attribute and its mutation with the additional group by attribute as the original query. 
Datasets can be generated using the techniques for killing mutations of additional group by attributes. 

 \subsection{Distinct Clause Mutations}
\label{sec:distinct}
Users may erroneously omit the DISTINCT keyword in the projection list of a select clause. 
For example, consider the following query from \cite{dbconcepts2010} that finds the department names of all instructors.
\\ \begin{small}
\verb|SELECT DISTINCT dept_name|
\verb|FROM instructor|
\end{small}

In this query, the absence of the DISTINCT keyword would lead to the same department name being repeated which is not desired. 
We term mutations that add or delete the DISTINCT keyword to the select as 
distinct mutations (DISTINCT of aggregates is covered in Section~\ref{sec:bg:mutation}).
To kill such mutations we need a dataset which  results in at least two tuples in the output such that these tuples  are identical on the projected attributes. 
\newstuff{We use the technique described in Section~\ref{sec:set:set_mutation} for generating tuples with identical projected attributes.}
For such a dataset, the query with the DISTINCT keyword will give only a single tuple as output while the one without, will give at least two tuples.

\newstuff{In case the constraints are not satisfiable, it is not possible to have multiple tuples with the same value of the projected attribute(s). 
This could happen if one of the projected attributes is a primary key for the input to the DISTINCT clause 
or if the projected attributes are also used as GROUP BY attributes in the same query block.
For such cases, the DISTINCT mutation is equivalent. 
}
 \section{Other Extensions}
\label{sec:misc}

\noindent\textbf{From clause subqueries}: Our parser turns from clause subqueries into a tree which can be handled using our existing techniques. 
We do not handle from clause subqueries that project aggregates, if there are constraints on the aggregation result in the enclosing query (other than simple constraints which our techniques handle) or if the query uses the lateral construct.  Handling such queries is an area of future work. \\

\noindent\textbf{Handling Parameterized Queries}:
When generating datasets for a query with parameters,  we assign a
variable to every parameter. The solution given by the SMT solver also contains a 
value for each parameter. 
It should be noted that since the solver assigns these values,
each dataset may potentially have its own values for the parameters. \\

\noindent\textbf{DATE and TIME}:
We handle SQL data types related to date and time,  
namely  DATE, TIME and TIMESTAMP by converting them to integers. \\

\noindent\textbf{Floating and Fixed Point Numbers}:
CVC3 allows real numbers to be represented as (arbitrary precision) rationals and hence 
when populating a real type data (floating or fixed precision)
from the database or query, we represent it as a fraction in CVC3.
When converting values to fixed precision values, supported by SQL, the conversion can in theory
cause problems in rare cases, since two rationals generated by CVC3
which are very close to each other may map to the same fixed precision number.
We have however not observed this in practice so far.\\

\noindent\textbf{BETWEEN operator}:
For queries that contain the BETWEEN operator, say \textit{attr BETWEEN a AND b}, we convert the BETWEEN operator to  $attr > a$ $AND$ $attr < b$. 
The datasets for killing selection mutations are also able to catch mutations where the user intended the range to include $a$ or $b$ or both. \\

\noindent\textbf{Insert/Delete/Update Queries}: 
To handle INSERT queries involving a subquery, and DELE\-TE queries,
we convert them to SELECT queries by replacing ``INSERT INTO relation''
or ``DELETE'' by ``SELECT *''.  
UPDATE queries are similarly converted by creating a SELECT query
whose projection list includes the primary key of the updated table, and the new values for
each updated column; the WHERE clause remains unchanged from the UPDATE query.
Data generation is then done to catch mutations of the resultant SELECT queries.

When testing queries in an application for correctness, we execute the original
INSERT, DELETE or UPDATE queries against the generated datasets.
To test student queries against a given correct query, we perform the transformation from
INSERT, DELETE and UPDATE queries to SELECT queries as above for both the given
student queries and the given correct queries, before comparing them
as described in Section~\ref{sec:tool}. \\

\noindent\textbf{Handling WITH Clause and Views}:
We syntactically convert a query using a WITH clause or views by
performing view expansion.  The assumptions we make about the query structure
must be satisfied by the resultant expanded query.\\

\noindent\textbf{ORDER BY clause}:
ORDER BY clause mutations include missing ORDER BY clause or attributes, additional ORDER BY clause or attributes, using ORDER BY DESC instead of ORDER BY and vice versa. 
In the absence of any ORDER BY clause, the order of tuples is determined by the query plan. 
Hence, it is possible for a query without an order by clause or with an incomplete order by clause, to give a result in the same order as a correct query, depending upon the chosen plan. Thus, order by mutations, in general, cannot  be caught by comparing results on different datasets, although we can use such comparison as a heuristic.
Mutations between ORDER BY and ORDER BY DESC can, however, be caught by generating appropriate datasets. 
To kill such a mutation we generate a dataset having two distinct values for the order by attributes.

As an alternative to checking results on generated datasets, 
mutations involving missing or additional ORDER BY clause or attributes can be detected by checking the ORDER BY clauses in the query. 
However, care should be taken to handle equivalent ORDER BY clauses due to functional dependencies, equality predicates between variables, and equality selection conditions. 

 \section{Grading Student SQL Queries}
\label{sec:tool}

In \cite{xdata:icde15} we describe the XDa-TA grading tool which uses 
datasets generated by the techniques presented in this paper for checking the correctness of student SQL queries.  
Here we describe how to efficiently check student queries against given correct queries. 
For each query in an assignment, a correct SQL query is given to the
tool, for which it generates datasets for killing mutants of that query.
To check if a student query is correct, the results of the student and correct query are
compared on each dataset.

It is to be noted that we do not aim to prove query equivalence of student query and correct query. 
Query equivalence between two queries $Q_1$ and $Q_2$ can be proven if we are able to prove that $Q_1$ is contained in $Q_2$ and \textit{vice versa}\footnote{Query containment can be reduced to equivalence similarly 
since $Q_1 \subseteq Q_2 \equiv Q_1 \cap Q_2 =Q_1$}. 
Thus, query equivalence can be modeled in terms of query containment.
Under set-semantics, it can be shown that the problem of query containment is NP-complete for conjunctive queries \cite{cm:stoc77}, 
and $\prod_2^p$-complete for queries involving inequalities \cite{ineq.equiv,conjuct.equiv}.  
For bag semantics, the complexity of query containment is undecidable for conjunctive queries with inequalities \cite{bag}.
We tried a sufficient condition for query equivalence,
namely that both $Q_i$ and $Q_{i,j}$ generate the same optimal query
plan, but as results in Section~\ref{sec:expt:grading} show, this
approach is often unable to establish equivalence of correct queries.

Thus, we only aim to catch common errors and it is possible that a non-equivalent student query may be marked correct.
However, in case we mark a student query as incorrect we have a dataset on which the student query and the correct query gives different results and hence guarantee that the queries are not equivalent.

The instructor needs to upload the schema and optionally small sample tables, by providing SQL script files. 
The instructor can then add assignment questions in text and correct queries for the same. 
For each correct query, the tool then generates datasets, 
using the techniques of the XData system. 
Each dataset is tagged with a label indicating what kind of mutation the dataset was designed to kill.
Student queries are submitted directly by the tool or can be uploaded in bulk. 

For some assignments, it may be possible to write correct queries using several very different approaches.
Datasets generated for a correct query are designed to be used to kill mutations of that query, 
but may or may not succeed in killing mutations of a different formulation of the query.
It could also happen that the question in text set by the instructor was ambiguous and there are multiple ways of interpreting it.
For these cases, the instructor mode allows multiple correct queries to be uploaded. 
Datasets generated from all the correct queries 
are used while evaluating student queries. 
The instructor may set whether datasets of all the queries need to be passed or only one query needs to be passed depending on the need. 
Besides, additional datasets for the query may also be added if desired. 


Let $Q_{i,j}$ denote the $j^{th}$ student's query submission for question $i$.
Let $CQ_{i,m}$ denote the $m^{th}$ correct query for question $i$ and $D_{i,m,k}$ be the $k^{th}$ dataset for the correct query $CQ_{i,m}$.

To evaluate student queries for a given correct query $CQ_{i,m}$, 
for each corresponding dataset $D_{i,m,k}$, the tool first uploads the dataset to the database, creating appropriate tables.
The tables created for this purpose are temporary tables whose view is limited for only a session so that there are no 
conflicts in case multiple student queries are being evaluated simultaneously. 
Next to compare the result of each student query $Q_{i,j}$
with that obtained by the correct query, $CQ_{i,m}$,
the tool executes an SQL query of the form \\
\begin{small}
($Q_{i,j}$ \texttt{EXCEPT ALL} $CQ_{i,m}$) \texttt{UNION} ($CQ_{i,m}$ \texttt{EXCEPT ALL} $Q_{i,j}$)  
\end{small}
on the temporary tables.

If the result of the above query is non-empty for any dataset $D_{i,m,k}$,
the student query $Q_{i,j}$ is marked as incorrect. 
If the results of the above query are empty
for {\em all} datasets, query $Q_{i,j}$ is deemed correct
for the purpose of grading.
The instructor can also decide that the presence of duplicates does not matter and in this case the tool uses EXCEPT instead of EXCEPT ALL in the query above.

An assignment can be marked as a learning assignment or a graded assignment.
When the tool is used in student mode, for 
graded assignments, the tool accepts queries from the student
and saves the queries for later grading.  Grading can be initiated
from by the instructor.
For learning assignments, the system executes the queries and
displays which datasets the query fails on (this can be done
incrementally, one failed dataset at a time).
Tagging datasets with the type of mutation that the dataset was
intended to kill, as mentioned earlier, helps students understand 
what the mistake was.

Our approach for checking the correctness of query relies on killing the mutations of the correct query and not of the student query. 
As a result, we may not catch erroneous student queries that have extra selection conditions. We do catch extra join conditions if the column names are identical but may miss other extra join conditions also.
Consider a query condition $x>3$. We generate datasets for satisfying $x>3$, $x=3$ and $x<3$. 
These datasets will catch the mutations involving a change in the operator and in case the condition is missing. \newstuff{
However, if the student query contains $x>3$ $AND$ $x<>2674$, it may be marked as correct since we may not have any test case to test mutation of $x<>2674$. 
Since the additional condition could be any arbitrary condition it is not feasible to generate datasets to catch such errors. 
One way to deal with this is to generate datasets based on mutations of the student query as well and use these also in grading. These datasets would catch such extra conditions. 
Since this requires a lot of overhead including constraint generation, constraint solving etc. for all the student queries, we do not implement this currently. 
We did not find any such student query in our experiment described in Section~\ref{sec:expt:grading}.}

\eat{
We also have a mutant generation tool, which can generate all
non-equivalent mutations (single {\em and} multiple) of a given query.
}


 \section{Related Work}
\label{sec:relwork}

The AGENDA toolset can generate test data for an application, given as input the database schema, the application
source code and certain sample value files. The data generated is however query agnostic, \newstuff{and may not catch errors if the selection conditions are not satisfied, leading to empty results in all cases.}
Reverse Query Processing (RQP) \cite{RQP} takes as input a query $Q$ and a result $O$, and
generates input data $I$ such that $O = Q(I)$, the result of $Q$ on input $I$. 
\newstuff{Since the query result needs to be provided as input, RQP cannot be used to test correctness of SQL queries.}

Qex \cite{SQE,QEX} is a tool for generating a dataset and 
parameter values for a given parameterized SQL query using the SMT solver Z3.
The goal is to generate data so that the query has a non-empty result. 
This corresponds to the generation of the first dataset in our case.
However, Qex does not address killing of query mutations.
\newstuff{Datasets of Qex may not be able to catch errors in join conditions, distinct, aggregate, missing or additional group by attributes as well as  missing selection or joins conditions. }

Tuya et al.~\cite{mutation1} describe a number of possible mutations for SQL queries. However, they do not handle test data generation for killing these mutations. 
They divide the mutations into four classes: mutations of the main SQL clauses (SC), mutations of the operators that are present in conditions and expressions (OR), mutations related to the handling of NULL values (NL), and replacement of identifiers: column references, constants and parameters (IR). We generate dataset for all of SC, OR and NL mutations except for the following: mutations related to arithmetic expressions, some mutations of LIKE patterns, mutations
between AND and OR, and some mutations related to three-valued logic. Currently, we do not consider IR mutations.
Handling the above mutations is an area of future work. 
However, we do consider some mutations that are not covered in \cite{mutation1} 
such as join type mutations on alternative join orders and mutations of the LIKE operator.

Riva et al.~\cite{Riva:2010} introduce rules which they call SQL Full predicate coverage (SQLFpc) rules,
which specify conditions that must be satisfied by test cases in order to kill each of a variety of
SQL query mutations; further rules to handle a larger class of SQL constructs and mutations
are described in their Web tool \cite{sqlfpc.web}.  However, they do not describe how to actually generate data. 
\cite{Tuya:2010} extends \cite{Riva:2010} by generating  constraints based on SQLFpc 
and solving the constraints using a constraint solver called Alloy \cite{alloy}.
However, it considers data generation and mutation killing for only numeric selection conditions and joins. 
Queries involving strings, aggregation, subqueries, group by and updates are not handled. 

Pan et al.~\cite{mutagen:ast13} describe \textit{Mutagen} which, given a database application, 
first generates program code mutants and SQL-query mutants by transforming constraints
from SQL queries to program code, and then uses PexMutator~\cite{pex} to generate data to kill the mutants. 
However, they only handle mutations of conditions in the where clause;
as far as we can tell from their brief description, the class of mutations they consider
is very small, and in particular, they do not handle aggregation, subqueries, join type mutations
set operators, distinct mutations and a number of other query features and mutations that we consider.

The work in this paper extends our earlier work on XData~\cite{xdata:icde11,xdata:dbtest13,xdata:icde15}; details of the differences and novel contributions of this paper were described earlier in Sections~\ref{sec:intro} and \ref{sec:bg}. 

Olston et al.\ \cite{olstonCS09} take a dataflow program and a database
and generate an example dataset such that the result of each operator 
(including intermediate operators) in the program is non-empty. 
However, they do not handle integrity constraints or check for query correctness.

There have been a number of papers for testing database applications. However, these do not address the problem of testing queries in the applications. 
Emmi et al.~\cite{emmi} and Pan et al.~\cite{mutagen:dbtest11,pan:2014} describe approaches to testing applications
based on creating  database states and test inputs, which can
ensure code coverage.  
Kapf\-hammer and Soffa \cite{kapfhammer} similarly consider test adequacy
of database driven applications.

 \section{Experimental Results}
\label{sec:expt}

We implemented the techniques for data generation described in this paper, as extensions to the XData system.
We show that our techniques for constrained aggregation (Section~\ref{sec:expt:cagg}) and subqueries (Section~\ref{sec:expt:subq}) are able to generate non-empty datasets and kill mutations in a number of cases.
In Section~\ref{sec:expt:tpch}  we show that our techniques are capable of generating datasets and killing mutations for the queries in the TPC-H benchmark.
In Section~\ref{sec:expt:grading} we evaluate our grading tool and show that it is better at catching student query errors than fixed datasets or correction by TAs. 

Each of the techniques we describe targets a different query construct or mutation and hence it does not make sense to compare the different techniques that we have proposed with each other.

\subsection{Constrained Aggregation}
\label{sec:expt:cagg}
In Section~\ref{sec:tupleAssgn} we described our approach for estimating the number of tuples for the purpose of data generation for queries containing constrained aggregation. 
In this section, we provide experimental results on the estimation of number of tuples per relation and subsequent data generation for a number of queries containing constrained aggregation.
The objective is to see if the tuple assignment technique (Section~\ref{sec:tupleAssgn}) assigns tuples in a manner that 
(a) can produce datasets to generate to non-empty result on the original query (this is the first dataset as mentioned in Section~\ref{sec:bg}) and (b) kill mutations related to the HAVING clause (aggregate mutation and comparison operator mutation of the HAVING clause).

For this experiment, queries which involve constraints on  aggregate operators along with one or more GROUP BY attributes were chosen. (The list of queries is provided in  \supplement{sec:app:expt}.)
Aggregates in both outer query block and subqueries are considered.
We also manually generated non-equivalent mutations by mutating the comparison operator (20 mutations) and aggregate operator (16 mutations) for the chosen queries, to test if the datasets could kill these mutations\footnote{\label{mutationgen}We do not use any automated tool to generate mutations. 
The mutations generated by an automated tool may or may not be equivalent to the original query. If our dataset fails to kill some of the mutations we would not be sure if that was because of the incompleteness of our tool or because of equivalence of mutation and the original query.}.

The results are shown in Table~\ref{tab:expt:consagg}.
For each constrained aggregation, the Tuples column shows the number of tuples assigned to each base relation. 
The columns Comparison Mutations and Aggregate Mutations show if all the non-equivalent mutations of comparison operator in HAVING clause and aggregate mutation respectively were killed by the generated datasets or not.
Query CA8 had two constrained aggregations, one in a subquery and one in the outer query block which are labeled as CA8a and CA8b respectively.

The datasets generated by XData was able to produce non-empty results on all queries. 
In terms of killing mutations, 35 out of the 36 mutations were killed. 
The mutation from MAX to MIN was not caught for Test Case CA2.
For killing mutation on MAX to MIN we need two distinct tuples, one which satisfies the aggregate constraint and one which does not.
Our tuple assignment method assigned only one tuple to the relation that had the MAX aggregate value and hence this mutation was not caught.
Handling such cases is an area of future work. 

\begin{table}
\begin{center}
\begin{small}
\tabcolsep=0.15cm
\begin{tabular}{|c|c|c|c|} \hline
 \textbf{Test} & \textbf{Tuples} &\textbf{Comparison} & \textbf{Aggregate}\\
 \textbf{Case}&& \textbf{Mutations} & \textbf{Mutations}\\ \hline
 CA1  &  1,2 &$\surd$& $\surd$\\ \hline
 CA2  &  1,1,2 &$\surd$& $\times$\\ \hline
 CA3  &  2,1,2&$\surd$& $\surd$ \\ \hline
 CA4  &  3,3,1,3 &$\surd$& $\surd$\\ \hline
 CA5  &  1,2,2 & $\surd$& $\surd$\\ \hline
 CA6  & 1,2,1,2,2 &$\surd$& $\surd$ \\ \hline
 CA7  & 1,1,3 &$\surd$& $\surd$\\ \hline
 CA8a  & 1,2,2 &$\surd$& $\surd$\\ \hline
 CA8b  & 1,2,2 &$\surd$& $\surd$\\ \hline
 CA9 &  5,5 &$\surd$& $\surd$\\ \hline
\end{tabular}
\end{small}
\caption{Tuple estimation for constrained aggregation}
\label{tab:expt:consagg}
\end{center}
\end{table}

\subsection{Subqueries}
\label{sec:expt:subq}
In Section~\ref{sec:subquery} we described various techniques for generating test data and killing mutations for queries containing where clause subqueries.
For this experiment, we chose queries involving various subquery connectives both with and without aggregates 
(The list of all queries is provided in  \supplement{sec:app:expt}) 
and check to see if XData is able to generate a dataset that produces non-empty result on the original query.
We also manually generated non-equivalent mutants by mutating the subquery connective (20 mutations) and the conditions in the subquery (20 mutations) to test if the datasets could kill these mutations\textsuperscript{\ref{mutationgen}}.

For all the queries considered XData could generate a dataset that produced non-empty result on the original query. 
The datasets generated by XData were able to kill all of the 40 query mutations that we considered.


\subsection{TPC-H queries}
\label{sec:expt:tpch}
We also tried generating test data and killing mutations for queries from the TPC-H benchmark.
\newstuff{We asked a few volunteers (who had not worked on the XData project) to generate 
specific types of query mutants. We tested to check if the datasets generated by XData 
could kill these mutations or not.
In case, XData was not able to kill the mutations we examined to check 
if the mutant was equivalent to the original query or not.
We only used non-equivalent mutants for measuring the performance of XData.}

Since our parser did not support certain query constructs we made minor changes (mainly syntactic) to the queries so that it could be parsed and the datasets could be generated. 
However, for checking whether the datasets generated a non-empty result or not, and for generation and killing of mutations we used the original queries. 

We were able to successfully generate datasets  for 17 out of the 22 queries. 
Of the 5 queries for which our techniques  failed to generate correct datasets, 4 queries had query constructs which are not currently handled (subqueries that have aggregates with expressions, aggregate value compared to a subquery and aggregate in a from clause subquery).
One query failed because the CVC3 solver crashed while generating datasets for that query.
Extending our system to handle these construct and migrating to newer version of CVC or using a different solver such as Z3 is an area of ongoing work.

The number of the different types of mutations killed across all queries is shown in Table~\ref{tab:expt:tpch}.
In addition to the mutations that our techniques explicitly target, we also tested queries with mutations of  arithmetic expressions (replacing one arithmetic operator with another).

\begin{table}
\centering
\begin{small}
\tabcolsep=0.15cm
\begin{tabular}{|p{3.5cm}|c|c|} \hline
 \multicolumn{1}{|c|}{\textbf{Mutation}}&\textbf{Mutants} &\textbf{Mutants}\\
 \multicolumn{1}{|c|}{\textbf{Type}} & \textbf{Generated}&\textbf{Killed}\\ \hline
 Selection (Comparison)& 10 & 10  \\ \hline
 Join Type (INNER / OUTER)& 8& 8  \\ \hline
 Aggregation (Distinct/ MIN vs.\ MAX)  & 9& 9 \\ \hline
 String Selection (String Comparison)& 7 & 7  \\ \hline
 String Like &5 & 5 \\ \hline
 Missing Joins Conditions &13 &13\\ \hline
 Having Clause (Comparison Operator) &2 & 2  \\ \hline
 Subquery Connective &6 &6  \\ \hline
 Changed Group By  &21& 20  \\ \hline
 AND vs.\ OR   & 16& 16 \\ \hline
 Arithmetic Operator &13 & 12  \\ \hline
 \textbf{Total} &\textbf{110}&  \textbf{108} \\ \hline
\end{tabular}
\end{small}
\caption{Number of mutants caught on TPC-H queries}
\label{tab:expt:tpch}
\end{table}

Overall XData was able to kill over 95\% of the non-equivalent mutants that we obtained.
For TPCH query 4, XData could not generate a dataset for killing extra group by attribute mutations and hence the corresponding mutation was not caught. 
Of the 13 queries with arithmetic operator mutations all but one were killed even though we do not explicitly target these mutations; explicitly targeting them is an area of future work.

\begin{figure*}
\begin{center}
\includegraphics[width=1\textwidth,keepaspectratio=true]{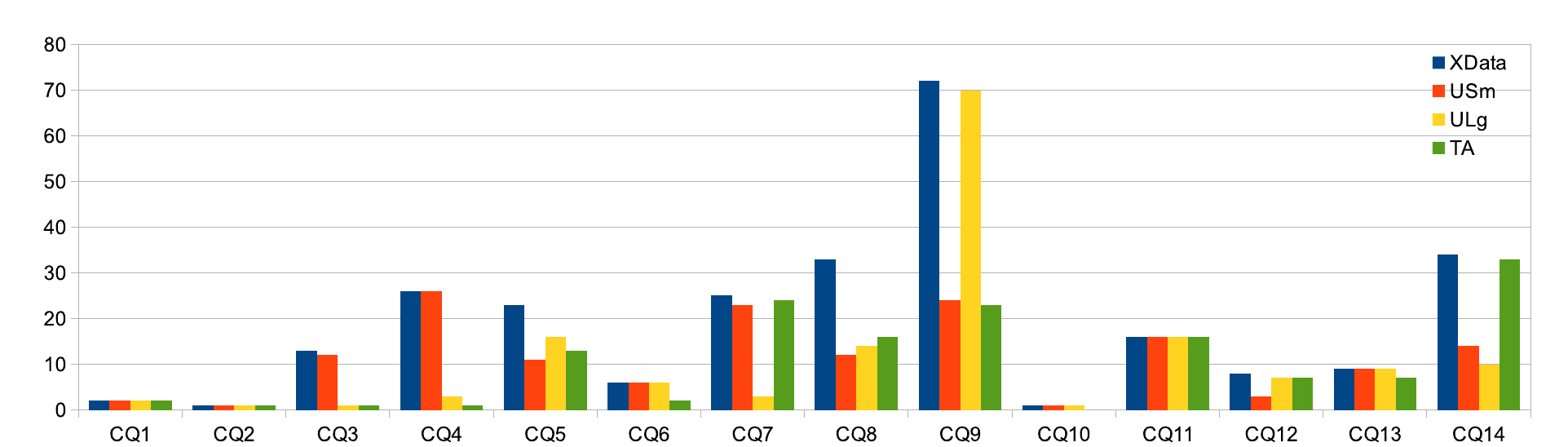}
\caption{Number of incorrect queries caught}
\label{fig:grading}
\end{center}
\end{figure*}

\subsection{Grading}
\label{sec:expt:grading}
We use the tool described in Section~\ref{sec:tool} to 
grade student queries. 
In order to compare grading done by XData to fixed datasets and the grading done by TAs, 
we used 14 SQL assignments, each of
which was answered by students of an undergraduate
database course at IIT Bombay.  
We omit questions which asked students to create DDL statements.

For each question, a correct SQL query $CQi$ was used to generate datasets.   The correct SQL queries 
are shown in \supplement{sec:app:expt}.
For the 9th assignment question, the query could be written in 2 quite 
different ways which we denote CQ9a and CQ9b; we generate datasets for 
both query formulations, and the results for CQ9 are using
the combined sets of datasets.
Query $CQ3$ was assigned at a point in the course where students had 
not been taught about the DISTINCT clause, and hence we set the 
testing tool parameters to ignore duplicates in the results of the
correct query and the student query.

The time taken for generating all the datasets for these queries 
(including the time taken by our code and the CVC3 solver)
ranged from 11 to 90 seconds, 
on a computer with an Intel(R) Core(TM) i5-2500K 3.30GHz CPU,
and 8 GB of memory, running Ubuntu.
\newstuff{The number of datasets generated ranged from 2 for CQ1 to 25 for CQ9a. 
Each dataset had a very small number of tuples, typically less than 5 per relation. 
The maximum number of tuples for a relation was 16.}


As comparison points, we also tested the queries with two sample University 
databases provided with the textbook by Silberschatz et al.\ \cite{dbconcepts2010},
and with the result of manual correction by course TAs.
The first University database, which we call USm is a small database which was
manually created by the authors of \cite{dbconcepts2010} to catch common errors; 
the second larger database, which we call
ULg is a larger database.
The TAs used a combination of testing against sample databases they
created, and their own reading of the queries. 

We also tried an alternate way to grade student queries,
by comparing the optimal query plans of the correct query with the 
optimal query plans for the student queries. 
If the plans match we flag the query as correct.  
We use PostgreSQL with the VERBOSE flag set to ensure that we get projected attributes of the query as well. 
Note that equivalent queries may not have identical plans.
For example, a condition $x>3$ is equivalent to $x>=2$ 
when $x$ is an integer, but plans using these alternatives would
be considered different. 
Also, the optimizer could find different plans for different ways of expressing the same query (especially true with subqueries). 
In our experiments we found that most of the student queries did not have the same plan as the correct query, even if they were correct (verified manually on sample cases).
For CQ3 the optimizer chose different join plans and hence most of the queries did not match. Same was the case with CQ7, CQ8, CQ13 and CQ14. \newstuff{For these queries, less than 5\% of the queries were marked correct.}

\newstuff{The result of the evaluations is shown in Figure~\ref{fig:grading}. 
Detailed evaluation is shown in Table~\ref{tab:query:eval} in \supplement{sec:app:expt}}.
For XData, USm and ULg the query is marked as incorrect iff there is a dataset that produces different results on the correct query and the student query.
Hence for these methods we can guarantee that a student query is marked incorrect only when it is not equivalent to the correct query. Consequently, the number of queries marked incorrect can be used as a measure of the effectiveness of the technique. 
\newstuff{We also tried to use the combination of USm and ULg grade queries. The number of incorrect queries caught turned out to be the maximum of the number of incorrect queries caught by USm and ULg.}

These results indicate that, overall, the datasets generated by 
XData were able to catch more incorrect queries than both USm and ULg, the
two  University datasets from \cite{dbconcepts2010}.
For CQ5, CQ8 and CQ14, in particular, our tool was significantly more efficient than the University datasets.

As compared to TAs, our datasets performed significantly better on many
queries, including CQ3, CQ4, CQ5, CQ6, CQ8 and CQ9.
The actual effectiveness of TAs is a little better than what the table
indicates, since there were some queries where students made minor
errors such as including extra attributes, which the
TAs decided to ignore as irrelevant, but which were caught by all 
the datasets.\footnote{If students had been told that their queries
would be graded by a tool, they would have probably taken more care to avoid
such errors.}

For query CQ5 and CQ8, some students had performed joins on the 
wrong tables, but these queries gave a correct result on 
datasets created by the TAs for checking the queries, and were
marked as correct.

For CQ8, the University dataset did not have 
any course taught by two different instructors in Spring 2010,
and hence a missing distinct keyword in the select clause was
not detected.  The TAs too did not enforce the check for distinct,
which was required for this query.

In contrast, for CQ4, the University dataset USm had a student who had taken CS-101
twice and hence performed as well as XData.
Again, the TAs had ignored the absence of a distinct specification. 
For CQ5, again the University datasets USm and ULg both
had some courses with two sections, which caught missing 
distinct specifications; in this case the TAs did check for 
the presence of the distinct specification.

For CQ9a a large  number of incorrect queries were caught by XData based on missing group by attributes and missing distinct clause.
For CQ14, the data generation and mutation killing technique for 
NOT EXISTS
was essential for catching a  large number of student query errors.



\paragraph{Discussion:}~\\
In order to get a measure of our accuracy or completeness of our techniques on these queries we need an oracle to identify which queries are correct and which are not.
This is very difficult for complex queries and doing this for classes with many students is extremely time-consuming.
The closest option is human evaluation.
However, our tool in its current version outperforms TAs (indicating TAs are not infallible). 
Hence, it is difficult for us to provide any completeness results for our grading tool. 

 \section{Conclusion}
In this paper we have addressed the issue of testing SQL queries and automated
testing of SQL student assignments. We used the XData system which we built earlier, to generate test datasets for
detecting errors, and realized that there were several limitations that needed to be addressed. We described several
novel extensions to address these limitations. We also tested
the efficacy of our test generation techniques for grading
SQL queries submitted by students, and showed that our
techniques outperform fixed (query independent) datasets,
as well as TAs in terms of catching errors, while avoiding
the drudgery of manual correction. Our XData system has great
potential for easing the life of database application developers and testers and also to database course instructors particularly to those of MOOCs. 

We have successfully  used  the grading tool in a UG database course at IIT Bombay to correct student queries.
The grading tool is available at \url{http://www.cse.iitb.ac.in/infolab/xdata} and  can be used by
course instructors for grading queries. \reminder{only mention here or add to intro/abstract}

Areas of future work include handling some SQL features
which we do not currently support, or support only partially,
and handling further classes of mutations.
These features include handling subqueries within a subquery, arithmetic expressions and mutations involving replacement of identifiers. Another area of future work is to award partial marks to student queries in a way that reflects how close the student query is to some correct query.

\begin{acknowledgements}
We would like to thank Tata Consultancy Services(TCS), India for partially funding this project through a grant and a PhD fellowship. 
We would also like to thank Amol Bhangadia, Bharath Radhakrishnan and Ankit Shah for their help in some running some experiments.
\end{acknowledgements}

\bibliographystyle{spmpsci}      

\bibliography{references}


\renewcommand\thesection{A\arabic{section}}

\newpage
\appendix
\normalsize

{\large{\textbf{Appendix}}}

\label{app}

\label{app}

\section{Cardinality Estimation for Join Inputs}
\label{sec:app:cardinality}
The tuple estimation for each relation for constrained aggregation on join result is done in 3 steps. First we construct a join graph. 
Then we infer attributes to be added to \textit{uniqueElements} and \textit{singleValuedAttributes}. 
In the third step, we assign cardinality to each relation such that the resulting number of tuples is $n$. 

\begin{figure}
  \centering
    \includegraphics[width=0.45\textwidth]{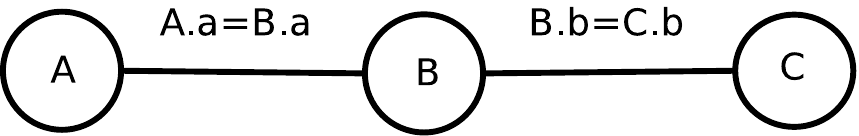}
       \caption{Join Graph }
       \label{fig:actsecond}
\end{figure}

\paragraph{Step 1: Construct Join Graph} ~\\
We construct a join graph \textit{G} = (\textit{R, E}), with
each relation in the query as a vertex. The join conditions from
one table to another are represented by a single edge between the
nodes. Figure~\ref{fig:actsecond} shows a join graph involving
relations $A$, $B$ and $C$. There are join conditions between $A$ and $B$, and
between $B$ and $C$. However there are no join conditions between $A$ and
$C$. Inferred join equalities are also added to the graph. For example,
the join conditions A.a = B.b and B.b = C.c imply that A.a = C.c is
also a join condition and hence it would be added to the graph. 
Note that this may introduce a cycle in the graph; 
our algorithm can work with cyclic join graphs.

\paragraph{Step 2: Infer Attribute Properties} ~\\
Next we apply the following sets of rules to infer properties of attributes

\textbf{Rule 1}: Every group by attribute is a single valued attribute.

\textbf{Rule 2}: Every set of attributes declared as primary key or unique key, is unique in the group. 

\textbf{Rule 3}: Every attribute which appears in conjuncts of the form \textsf{A.a=constant} is a single valued attribute.

\textbf{Rule 4}: If each attribute of any \textit{uniqueElements}[$R_{i}$] is a single valued attribute then all attributes of that relation are single valued attributes.

\textbf{Rule 5}: If any attribute, \textsf{$R_{i}.x$}, is a single valued attribute then every attribute of equivalence class (Section~\ref{sec:bg:mutation}) in which \textsf{$R_{i}.x$} is present becomes a single valued attribute. For example, if the join condition is A.a = B.a and A.a is single valued, B.a also becomes single valued. 

\textbf{Rule 6}: If an attribute of a unique element is single valued
then remaining attributes of unique element become unique. We apply
this rule recursively on  the unique element to get a minimal unique element. 
We then drop all non-minimal sets from $uniqueElements$.
For example, if (A.a, A.b, A.c) is unique and A.a is single valued then (A.b, A.c) is unique
and is added to $uniqueElements[R_i]$.  In this case (A.a, A.b, A.c) is dropped
from $uniqueElements[R_i]$.

\begin{algorithm}
 \renewcommand{\algorithmicrequire}{\textbf{Inputs:}}
    \renewcommand{\algorithmicensure}{\textbf{Output:}}
\begin{algorithmic}[1]
\REQUIRE \textit{joinConds}[$R_{i}$, $R_{j}$]\eat{:join conditions $\forall R_{i}, R_{j}$ $\in$ \textit{R}}\\ \textit{groupByAttributes}\eat{:group by attributes of the query}\\ \textit{unique keys} and \textit{primary keys} of relations $\in$ query
\ENSURE Effective \textit{uniqueElements}, \textit{singleValuedAttributes} for the query            
	
\STATE Build equivalence classes   using \textit{joinConds}[$R_{i}, R_{j}$], $\forall R_{i}, R_{j}$ $\in$ $R$

\STATE Apply \textit{\textbf{Rule 2}} and update \textit{uniqueElements}
\STATE Apply \textit{\textbf{Rule 1}}, \textit{\textbf{Rule 3}} and update \textit{singleValuedAttributes}

\WHILE{change in \textit{uniqueElements} \textbf{or}  \textit{singleValuedAttributes}}

	\STATE Apply \textit{\textbf{Rule 4}}, \textit{\textbf{Rule 5}} and update \textit{singleValuedAttributes}
	\STATE Apply \textit{\textbf{Rule 6}} and update \textit{uniqueElements}
\ENDWHILE
\RETURN{\textit{uniqueElements}, \textit{singleValuedAttributes}}

\end{algorithmic}
\caption{ : getAttributeInferences()}
\label{algo:getAttributeInferences}
\end{algorithm}

The rules are applied according to Algorithm~\ref{algo:getAttributeInferences} to infer which attributes are added to \textit{uniqueElements} and which to \textit{singleValuedAttributes}.\\

\paragraph{Step 3: Assign Cardinality} ~\\
\eat{A naive approach to assigning cardinalities is to assign $n$ tuples to each relation. 
However if some join attribute is not unique in both the participating relations, more than $n$ tuples could be present in the join result. }

We define some more terms
\begin{itemize} \itemsep 0em
\item $joinAttributes[R_{i},R_{j}]$: attributes of relation $R_{i}$ that are involved in join conditions with relation $R_{j}$.
\item $unique[R_{i},R_{j}]$: $\{ S_k \mid S_k \subseteq \mbox{joinAttributes}[R_i,R_j] \wedge S_k \in \mbox{uniqueElements}[R_i] \}$.
\item $n_{R_i}$: number of tuples assigned to relation $R_i$.
\end{itemize}


In order to find the number of tuples for each relation we use the attributes inferred using Algorithm~\ref{algo:getAttributeInferences} along with the following rules.

\textbf{Rule 7}: If $n_{R_{i}}$=$n,n>1$ and \textit{unique[$R_{i}$, $R_{j}$]}$\neq \emptyset$  then $n_{R_{j}}$ is set to $n$.
We also infer further unique elements as follows.
For each $S_k \in \textit{unique}[R_i, R_j]$, let $S'_k$ be the attributes
from $R_j$ that are equated to $S_k$.  Then add $S'_k$ to 
\textit{uniqueElements}[$R_j$].


The intuition behind Rule 7 is as follows. 
Consider the join of two relations A and B. Let the join condition be 
$A.a=B.a$ and suppose that $\{A.a\} \in$ \textit{uniqueElements}$[A]$. 
Here \textit{joinAttributes} [$A$, $B$]=\{A.a\}, 
\textit{joinAttributes} $[B, A]$=\{$B.a$\}, 
\textit{unique} [$A$, $B$]=\{$A.a$\} and 
\textit{unique} [$B$, $A$]=$\emptyset$.
If the cardinality of 
$A$ is $n$, since $A.a$ is unique, it must have $n$ different 
values. 
The relation $B$ has join condition with $A.a$ which belongs to 
\textit{uniqueElements[A]}. 
So $B$ must contain $n$ tuples with distinct values for the attribute
$B.a$ across $n$ tuples and each value matches with the value of 
$A.a$ for one of the tuples in $R_i$. 
So the cardinality of $B$ become $n$ and $B.a$ becomes a unique attribute.

\textbf{Implementation Rule 1}: 
If $n_{R_{i}}$=$n,n>1$ and $R_{i}$ has a multi attribute unique element, $mu$, such that every attribute of $mu$ participates in some join conditions but \textit{joinAttributes} $[R_i,R_j]\subset$ $mu$ for all j, then for at least one relation $R_k$ that joins with $R_i$ \textit{joinAttributes} $[R_i,R_k]$ is unique and  $n_{R_k}=n$. One such $R_k$ is picked and
we add \textit{joinAttributes} $[R_i,R_k]$ to \textit{uniqueElements}$[R_i]$ and \textit{joinAttributes} $[R_k,R_i]$ to \textit{uniqueElemen\-ts[$R_k$]}.


The intuition is as follows.  Consider the join graph shown in Figure~\ref{fig:actsecond}. Let joinConds[$A$, $B$]=\{A.a=B.a\}, joinConds[$B$, $C$]=\{B.b=C.b\}. Let (B.a, B.b) be unique. 
Here, \textit{joinAttributes} [$A$, $B$] = \{A.a\}, \textit{joinAttributes} [$B$, $A$] = \{$B.a$\}, \textit{joinAttributes} [$B$, $C$]=\{$B.b$\}, and \textit{joinAttributes} [$C$, $B$]=\{$C.b$\}. Further, \textit{unique} [$A$, $B$]=$\emptyset$, \textit{unique} [$B$, $A$]=$\emptyset$, \textit{unique} [$B$, $C$]=$\emptyset$, \textit{unique} [$C$, $B$]=$\emptyset$.


Suppose cardinality of $B$ is $n$. 
Since \textit{unique}[$B$, $A$] = $\emptyset$, is possible that $n_{A}$ =1 such that $A.a$ matches with all values of $B.a$ across $n$ tuples. 
Here $B.a$ contains same value across $n$ tuples. Similarly, we can choose $n_{C}$ = 1 and $B.b$ will have the same across $n$ tuples. 
Now both $B.a$ and $B.b$ have same values across all $n$ tuples. But (B.a, B.b) must be unique across $n$ tuples. 
So the assignment of cardinalities is incorrect. Hence at least one of B.a or B.b must be chosen to be unique, and this will cause one $n_A$ or $n_B$ to be $n$. 

Note that in this example had (B.a, B.b, B.c) been unique, every attribute of $mu$ does not participate in any of the join conditions. 
In this case, the rule is not applicable and both $A$ and $C$ may have a cardinality of 1. 
To generate $n$ tuples for $B$ such that the join results in $n$ tuples, B.c can have $n$ distinct values while B.a and B.b have same values corresponding to A.a, C.b respectively.

We differentiate this rule from others since this rule can have several possible outcomes as opposed to the other rules for which the outcome is definite and unique. 
One outcome is chosen. 
The choice of which of the joining relations is assigned cardinality as $n$ can be made by the solver or as heuristic the choice can be made arbitrarily; we describe these below.

\paragraph{Cardinality Inference Algorithm} ~\\
Let the aggregated attribute be $R.a$. 
For getting the cardinality of each relation, using the rules and the given join conditions of the relations we can encode the tuple assignment problem in the form of constraints in CVC3.
We add the following constraints in CVC3. 
\begin{itemize}
 \item constraints ascertaining \textit{singleValuedAttributes} and \textit{uniqu\-eElements} for each relation
 \item for each relation such that all attributes are single valued (Rule 4) constraints to ensure that the number of tuples is 1
 \item constraints for Rule 7 and the Implementation Rule 1 for all the relations in the query as applicable
 \item constraints to ensure that the final count after joining the tables is n
 \item in case n values are required for some attribute R.a to satisfy some aggregate condition we add constraints to ensure that the relation R has n tuples. 
 For example, consider a case where $SUM(R.a) = 17$, where $a$ is an integer attribute  and there is a constraint $R.a\leq 5$, we need at least 4 tuples for the given group of R and they cannot all be the same.
 It is not possible to satisfy the aggregation condition if we assign a single tuple to R, the join of R with other relations produces 4 tuples for the group.
 Similar is the case with SUM DISTINCT on an integer attribute. 
 
\end{itemize}

On solving this set of constraints, we get the number of tuples for each relation.

The constraint approach for tuple generation works well if the number of attributes is not very large. 
In practice, we use a simple and fast heuristic approach described as follows.
If any non-empty set of attributes of a relation forms a unique element and every attribute of that unique element is a single valued attribute then that relation must contain a single tuple (explained in Rule 4). 
For such relations, the only possible choice of cardinality is $1$. 
Of the remaining relations, the heuristic algorithm chooses one relation and assigns to it a cardinality of $n$, making it the root node. 
The count of all other nodes of the join graph, $n_{R_{i}}$ is initialized as 1. 
The root node ($R_{r}$) is then used as a starting relation to calculate the actual cardinality for other relations using Rule 7 and Implementation Rule 1.
The procedure for this is described Algorithm~\ref{algo:actualCardinality}.
If the heuristic fails we use the constraint approach. 

\begin{algorithm}
  \renewcommand{\algorithmicrequire}{\textbf{Inputs:}}
    \renewcommand{\algorithmicensure}{\textbf{Output:}}
\begin{algorithmic}[1]
\REQUIRE $G$ = ($R$, $E$): Join graph \\ \textit{singleValuedAttributes} \\  \textit{uniqueElements} \\  $R_{s}$: relation chosen as root node (cardinality can be $n$)

\ENSURE assigned cardinality $n_{R_{i}}$, $\forall R_{i}$ $\in$ $R$

\STATE $\forall R_{i}$ $\in$ $R$, initialize  $n_{R_{i}} \leftarrow 1$
\STATE nQueue = $\emptyset$
\STATE $n_{R_{s}}$ = $n$. 
\STATE nQueue.enqueue($R_{s}$)

\WHILE{$nQueue \neq \emptyset$ $R_i \leftarrow $nQueue.dequeue()}
	\FOR{each edge $E_{k}$ $\in$ $E$ from $R_{i}$ to $R_{j}$}
			\STATE prevCardinality $\leftarrow n_{R_j}$
			\STATE Apply \textit{\textbf{Rule 7}} from $R_{i}$ to $R_{j}$.
			\IF {change in $uniqueElements[R_j]$ or \\ (prevCardinality=1 and $n_{R_i}=n$)}
			  \STATE nQueue.enqueue($R_{j}$)
			\ENDIF		 
	\ENDFOR
	\IF{Implementation Rule is applicable on $R_i$}
	  \STATE Apply Implementation Rule 1 on $R_i$, let $R_k$ be the relation for which cardinality if to be changed to $n$
	  \STATE prevCardinality $\leftarrow n_{R_k}$
	  \STATE $n_{R_k} \leftarrow n$
	  \IF {change in $uniqueElements[R_k]$ or \\ prevCardinality=1}
	    \STATE nQueue.enqueue($R_{k}$)
	  \ENDIF
	  \IF {change in $uniqueElements[R_i]$ on Rule 8}
	    \STATE nQueue.enqueue($R_{i}$)
	  \ENDIF
	\ENDIF
\ENDWHILE
\RETURN{$n_{R_{i}}$, $\forall R_{i}$ $\in$ $R$}
\end{algorithmic}
\caption{ : getActualCardinalityHeuristic()}
\label{algo:actualCardinality}
\end{algorithm}

\section{Solving String Constraints}
\label{sec:app:string}
In this section, we describe our techniques to solve
string constraints. 
We also show some more experimental results comparing our string solver to other available string solvers.

\subsection{String Solver}
\label{sec:app:ssolver}
In this section we describe the working of our string solver.
To illustrate our method we use the following set of constraints as an example

\begin{exmp} 
\label{ex:string}
\begin{verbatim}

A > B
A like `%pqr%'
B ilike `_abc'
C >= B
C = `Biology'
A = E
E like `%abc%'
F >= B
G like `Bio%'
\end{verbatim}
\end{exmp}
In this example for the purpose of simplicity of representation we consider that the strings may take only alphabetical values.

Our solver works as follows.

\paragraph{Step 1: Collect Conditions.}~\\ From all the constraints required for generating a dataset for the query, in the first step, we separate and collect
the string constraints, i.e., selection conditions on strings,
like conditions, and string length conditions.

\paragraph{Step 2: Reduce Number of Conditions.}~\\
Next, we reduce the number of string constraints by removing the conditions containing the \textit{equality} operator as follows:

a) For each condition of the kind $S_i$ $=$ $const_i$, where $S_i$ is a
string variable and $const_i$ is a constant, we replace all occurrences of $S_i$
with $const_i$. This may lead to constraints of the form $const_i$
\relop\ $const_j$ or $const_i$ \likeop\ $pattern$. Using string operations, we then verify if such constraints  are satisfiable. 
If they are satisfiable then we remove the equality conditions 
else we infer that there is no possible solution to the given set of conditions. 
For example, if the conditions are $A=$`Comp' and $A$ \textit{LIKE}
`Bio\%', replacing the value of $A$ as `Comp' in
the latter condition leads to an unsatisfiable
 constraint. 

b) For constraints of the form $S_{i}$ =
$S_{j}$, we replace all occurrences of $S_{i}$ by $S_{j}$ 
in all constraints and
remove the constraint $S_i = S_j$ from the set. When an instance of $S_{j}$ has been found, after solving the rest of the constraints, we assign the same
value to $S_{i}$. 

In Example~\ref{ex:string} we assign C= `Biology' and replace all occurrences of C with this value. Replacing C, in C$>=$B we get `Biology'$>=$B. We rewrite this as B$<=$`Biology'. Since A=E is a constraint we replace all occurrences of A by E. 
After this step the constraints are 

\begin{figure}[b]
\begin{center}
\includegraphics[width=0.15\textwidth,keepaspectratio=true]{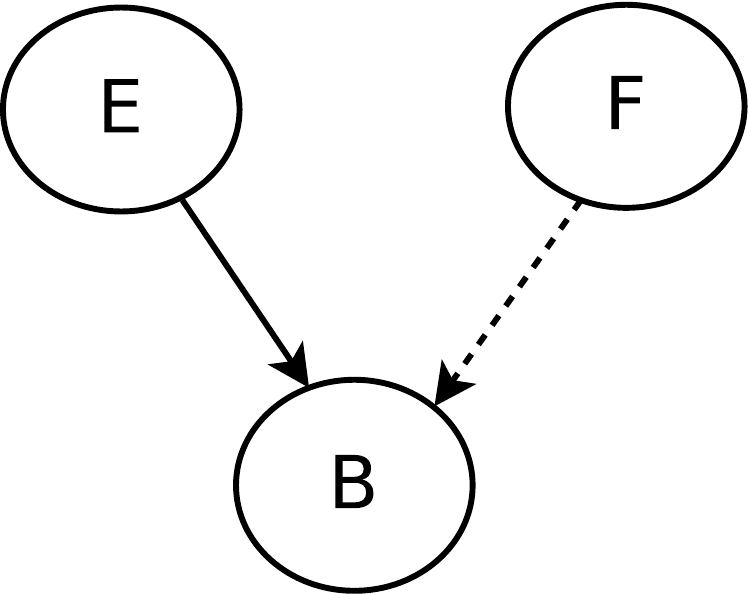}
\caption{Dependency among string variables}
\label{fig:string.dep}
\end{center}
\end{figure}
\begin{verbatim}
E > B
E like `%pqr%'
E ilike `_abc'
B <= `Biology'
E like `%abc%'
F >= B
G like `Bio%'
\end{verbatim}

\paragraph{Step 3: Group Related Variables.}~\\
Next, we group variables that depend on each other, i.e., if
$S_{i}$ \relop\ $S_{j}$ or $S_{i}$ \likeop\ $S_{j}$ is present in the set of constraints then $S_i$ and $S_j$ are in the same group. 
Once these groups are formed, we then solve the constraints for one group at a time. This grouping of variables helps in reducing the number of constraints that need to be solved at a time.
In the above example E, B and F are dependent on one another and are hence grouped together in a group  while G is put in another group.

For each group,
we construct a graph, where the variables form the vertices. Let vertex $V_i$ represent the string variable $S_i$. A constraint of
the form $S_i < S_j$ or $S_i \le S_j$ is represented by a
directed edge from $V_j$ to $V_i$ in the graph.
Constraint $S_i <> S_j$ is
represented by an undirected edge between $V_i$ and $V_j$. 

The graph for our example case would look like the one shown in Fig.~\ref{fig:string.dep}
where the dotted edge between F and B implies $\leq$ and the edge between E and B implies $<$.

Additionally, for each string variable, $S_i$, we store the following information.
\begin{itemize} \itemsep 0em
 \item \textit{MaxLength}: The maximum allowable
length of the string.  It is initially assigned a default value. This
value is modified based on string length constraints on $S_i$, if any.

\item \textit{MinLength}: The minimum allowable length of the string. Similar to the \textit{MaxLength} this also has a default value and is modified based on length constraints. 

\item \textit{NotEqualLengths}: This is a set of values
of length values not allowed for $S_i$. This captures constraints of
the kind $strlen(S_i) <> $ \textit{constant}.

\item \textit{Less}: list of variables with the value \textit{less} than $S_i$
\item \textit{LessEqual}: list of variables with the value is \textit{less} than or equal to  $S_i$.

\item \textit{NotEqual}: list of variables with the value not equal to $S_i$.

\item \textit{OtherConstraints}: This list contains constraints
of the form $S_i$ \relop\ \textit{constant} or $S_i$ \likeop\
\textit{pattern}.

\end{itemize}

\paragraph{Step 4: Choose the Variables to Solve.} ~\\
We traverse the
graph and first collect all vertices $V_1,..,V_k$ whose outdegree is
$0$. These vertices represent the string variables whose value is the lowest amongst
all comparable variables.
In our example we choose the variable B. 


If we do not find any such variable, it implies that there is a cyclic
dependency among variables with each variable being  less than (equal) to that
some other variables.  Essentially, this means that either all the
variables in that cycle are equal to each other, if all edges are $\leq$,
or that the given set of constraints is not satisfiable, if at least one of the edges is $<$. 
We first solve for these variables (with outdegree 0), one by one, using the function  \textit{Solve\-One\-Variable} (described below) which finds the lexicographically smallest string possible.

After  obtaining the solution for a vertex, say $V_i$ (and hence string variable $S_i$),
 for each vertex  $V_j$ (string $S_j$) that has an edge to $V_i$ in the graph, 
we add appropriate constraints, using the solution of $S_i$ to the list of constraints for $S_j$. 
 We then remove $V_i$ from the graph and solve for the remaining vertices by repeating this step on the modified graph. 


We now describe the function \textit{SolveOneVariable} for finding the
solution for vertex, $V_i$.
This function consists of two parts a)
building an automaton and b) finding the  lexicographically smallest string on
this automation that satisfies all the constraints.

\textit{Step 4a: Building an automaton}: 
We first convert the constraints of the form $S_i$ \relop\ constant and $S_i$ \likeop\ pattern to $S_i$ \textit{matches} $re$ where $re$ is the corresponding regular expression in Java. 
This conversion is made by functions written
specifically for each LIKE and comparison operator, as illustrated by the 
following examples of conversion:

\begin{small}
S1 $>$ `Bio'  $\rightarrow$ \verb$`[C-z]\w*|B[j-z]\w*|Bi[p-z]\w*|Bio\w+'$ 

S1 LIKE `Bio\%' $\rightarrow$ \verb$`Bio\w*'$ 

S1 LIKE `Bio\_' $\rightarrow$ \verb$`Bio\w'$  

S1 ILIKE `Bio\%' $\rightarrow$ \verb$`[B|b][I|i][O|o]\w*'$ 
\end{small}
\\(\verb$\w$ denotes a wild character)

We build an automaton, $A$ for the identity pattern (`\textbackslash w*'). Then for every constraint
that must be satisfied by $S_i$, we create another automaton, $B$ and modify
the automaton $A :=  A \cap B$. We use a slightly modified version of the
automaton package dk.bricks.automaton \cite{AUTO} operation on automata.
We use our own methods for converting a given Java compatible regular expression to an automaton. 
If the number of constraints on a variable is above a certain threshold we minimize the automaton resulting from $A \cap B$ at each step so as to improve the performance. 

\eat{After all the constraints have been added to the automaton, we convert the
automaton to a deterministic automaton and minimize it. For this purpose we use
the functions provided in the automaton package itself.}

\begin{algorithm}
 \caption{getSmallestString (state, strBuilder, max, min, notEqualLength, length)}
\label{SM}
\begin{algorithmic}[1]
 \IF{length$>=$ min $\wedge$ length$<=$max $\wedge$ state $\in$ finalState $\wedge$ length $\notin$ notEqualLength}
\STATE Return strBuilder
\ENDIF
\IF{length$>$max}
\STATE Return null
\ELSE
\STATE  tran = state.getTransitionsSorted
\FOR{$\forall$ c $\in$ tran}
\STATE append c to strBuilder
\COMMENT{Each transition is on a character}
\STATE str= getSmallestString(c.to, strBuilder, max, min, notEqualLength, length+1)
\IF{str $\neq$ null}
\STATE Return str
\ENDIF
\STATE remove last char from strBuilder
\ENDFOR
\ENDIF
\STATE Return null
\end{algorithmic}
\label{algo:string}
\end{algorithm}

\textit{Step 4b: Finding the lexicographically smallest string} :
Once we have the minimized automaton, $A$, for a variable, $S_i$,
we find the lexicographically smallest possible string  within
\textit{MaxLength} and \textit{MinLength} for that $S_i$.
To find such a string, we use a backtracking approach which traverses
the automaton graph in a depth-first manner.
At each step we check if (a) the current depth $>=$ \textit{MinLength}
and $<=$\textit{MaxLength}, (b) the state is a final state, (c) the current depth is not present in NotEqualLengths. 
If these conditions are satisfied then we return the string obtained by the traversal.  If these conditions
are not satisfied then even after reaching the dept of \textit{MaxLength},
we backtrack. 
If after traversing the entire graph, we do not find a string 
that satisfies the conditions then we return a null value.
Details are provided in Algorithm~\ref{algo:string}.

For our example, an automaton is created for B using the constraints on B i.e B$<$`Biology'. 
The smallest possible value for B is found to be `A'. 
We then add the constraint E$>$`A' to E and F$>=$`A' to F and remove B from \textit{E.Less} and \textit{F.LessEqual}. 
Now the remaining variables E and F do not have any dependency on each other and can be solved in any order. 
We create appropriate automata for both the variables and find suitable values using Algorithm~\ref{algo:string}.
Now in order to satisfy the condition A=E after solving the variables B, E and F we put the value of A the same as the one obtained for E.

\paragraph{Constraints containing ``$<>$'' and ``$\sim =$''}:\\
We handle conditions of the kind $S_i \sim = S_j$  
and $S_i <> S_j$, where both $S_i$ and $S_j$ are string variables,  such that one of $S_i$ and $S_j$ is unconstrained, i.e., there
are no other string constraints constraining the value of one of them.
For such cases, we first find an assignment to the constrained
variable and then assign a value of other variables that satisfies the $<>$ or $\sim =$ constraint as applicable.

\subsection{String Solver Performance}
\label{sec:app:string:expt}

The experiment in this section focuses on the performance of our string solver as compared to other solvers in terms of the time taken to solve string constraints.  
The experiments for HAMPI \cite{HAMPI}, Kaluza \cite{KAL}, CVC4 \cite{CVC4}, SUSHI \cite{SUSHI} and XData string solver were run on a virtual machine with 4GB RAM and a dual core CPU running Ubuntu Linux. For Rex \cite{REX} we used a  virtual machine with the same configuration running Windows~7.

\begin{table}
\begin{center}

\tabcolsep=0.15cm
\begin{tabular}{|c|p{4.4cm}|} \hline
 
 \bf{Test} & \multirow{2}{*}{\bf{Constraints}} \\
 \bf{Case} & \\ \hline
 S1 &  A like `Comp\_\_' \\ \hline
 S2 &  A like `Mr\%' \\ \hline
 S3 &  A ilike `\%sr\%' \\ \hline
 S4 &  A like `Comp\%', A like `\%Sc'\\ \hline
 S5 &  A like `Comp\%', A like `\_Sc'\\ \hline
S6 &  A $>$ `Bio' \\ \hline
 S7 &  A like `\%Sc', A like `Life\%', A.length $>$ 6\\ \hline
  S8 & A $<$ B, B like `Bio\%', \mbox{A like `CSE\%'}  \\ \hline
 S9 &  A$<$B, B like `Bio\%', B.length$>$4, A like `\%101' \\ \hline
 S10 &  A $>$ B, A like `\%pqr\%', B ilike `\_abc', C $>=$ B, \mbox{C = `Biology'}, A = E, \mbox{E like `\%abc\%'}, F $>=$ B, \mbox{G like `Bio\%'} \\ \hline
\end{tabular}

\caption{String solver test cases}
\label{tab:expt:const}
\end{center}
\end{table}

For the first experiment, we study the efficiency of the string solvers in a variety of common cases. 
The test cases for this experiment are listed in Table~\ref{tab:expt:const}. 
We include a mix of satisfiable and unsatisfiable cases. 
The last 3 test cases contain multiple string variables and can only be solved with our string solver. 
We include these cases to show that the performance does not drop much even when solving for multiple variables. 
For solvers other than the XData string solver the expressions in the form of A \textit{likeop/relop} \textit{expr} etc.\ 
were manually converted to regular expressions of the format recognized by the solvers.
The running time does not take into account the conversion. 
For XData string solver we fed constraints in the same form as in the SQL queries and let XData convert these to regular expressions. 

The time taken by different string solvers for this experiment is shown in Table~\ref{tab:expt:string}. 
The test cases that cannot be solved by a particular solver\footnote{HAMPI currently has a known bug because of which it cannot handle more than one constraints on the same variable in some cases. Test cases 4, 5 and 7 failed because of this.} is marked with a ``-'' and cases that ran for a very long time ($>$20 min) but still did not terminate are marked with a ``*''. 
In terms of time taken, CVC4 and the XData solver turn out to the most efficient ones for these cases, but CVC4 cannot handle comparison among multiple variables. \footnote{\newstuff{We tried to encode string comparison as user defined functions in CVC4 but with these functions the execution did not terminate even after 20 min.}}

\begin{table}
\begin{center}
\begin{small}
\tabcolsep=0.15cm
\begin{tabular}{|c|c|c|c|c|c|c|} \hline
 
 Test & \multirow{2}{*}{HAMPI} & \multirow{2}{*}{Kaluza} & \multirow{2}{*}{SUSHI} &\multirow{2}{*}{CVC4} & \multirow{2}{*}{Rex} & XData \\ 
 Case & & & & & & solver \\ \hline
 S1 & 150 	& 706 	& 22	&6	& 124 	&4\\ \hline
 S2 & 136 	& 706 	& 34	&6	& 140 	&4 \\ \hline
 S3 & 139 	& 708 	& 39	&9	& 140 	&4\\ \hline
 S4 &  - 	& 2444 	& 175	&17	& 168	&15\\ \hline
 S5 &  - 	& 671 	& 160	&19	& 156 	&14\\ \hline
 S6 & 137 	& 380	& 54	&*	& 256	&4\\ \hline
 S7 &  - 	& 653	& -	&20	& - 	&11\\ \hline
 S8 &  - 	& - 	& -	&-	& -	&23\\ \hline
 S9 &  - 	& -	& -	&-	& - 	&11\\ \hline
 S10 & - 	& -	& -	&-	& - 	&30\\ \hline
\end{tabular}
\end{small}
\caption{Time taken by string solvers (in ms)}
\label{tab:expt:string}
\end{center}
\end{table}

We conducted two experiments to test the scalability of the solvers. Scalability can be measured in terms of length of string that can be successfully solved by the solver or by the number of simultaneous constraints it can handle.

\begin{figure}
\begin{center}
\includegraphics[width=0.45\textwidth,keepaspectratio=true]{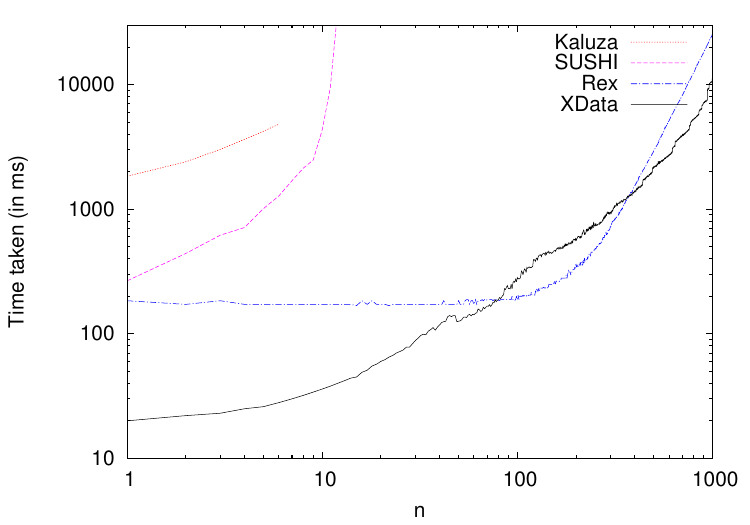}
\caption{Time taken vs.\ length of result}
\label{fig:string.scale}
\end{center}
\end{figure}

For the second experiment, we use the experimental benchmark from Rex \cite{REX} to measure the performance as the length of the string required in the output varies. 
The constraint to be satisfied by the string is that must match intersection of regular expressions \\ \textbackslash $w*[a-c]*a[a-c]\{n+1\}$\textbackslash $w*$ \\and\\ \textbackslash $w*[a-c]*b[a-c]\{n\}$\textbackslash $w*$ \\ $n$ is a parameter which we varied from $0$ to $1000$. The results for this experiment are shown in Fig~\ref{fig:string.scale}.

CVC4 and HAMPI failed to generate any result for any value of  $n$ and hence could not be included. 
KALUZA gave the result as UNSAT (cannot be satisfied) for $n>6$ while SUSHI ran out of memory for $n>13$.
Rex and XData solver were able to successfully generate string till $n=1000$.
In terms of time taken, the XData string solver turned out to be the most efficient for most cases.

\begin{figure}
\begin{center}
\includegraphics[width=0.45\textwidth,keepaspectratio=true]{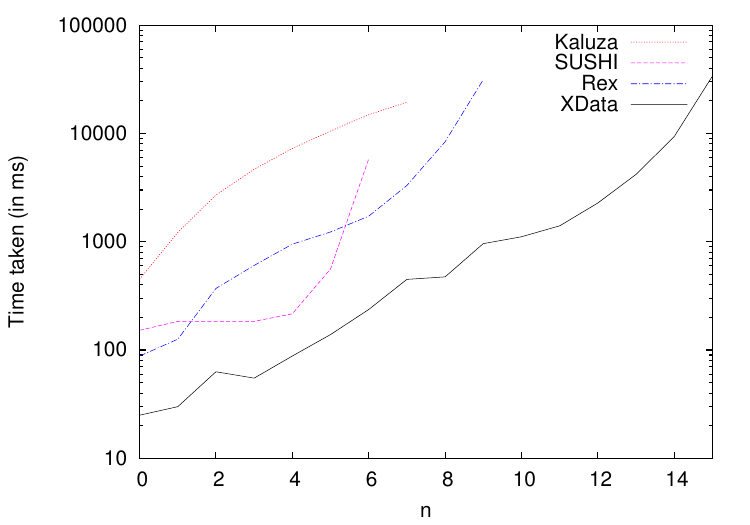}
\caption{Time taken vs.\ number of constraints}
\label{fig:string.stress}
\end{center}
\end{figure}

For the third experiment, we measure the performance in terms of time taken to solve varying number of constraints. 
For each $n$ the constraint to be satisfied is that the string must match the intersection of regular expressions \textbackslash $w*[a-c]*b[a-c]\{i\}$\textbackslash $w*$, $\forall i$, $0\leq i \leq n$. 
We varied $n$ from $0$ to $15$. The  results are shown in Fig~\ref{fig:string.stress}.

Here again CVC4 and HAMPI failed to generate any result any value of  $n>0$ and hence could not be included.
KALUZA gave UNSAT result for $n>7$.
SUSHI and Rex ran out of memory for $n>9$ and $n>6$ respectively.
In this experiment also the XData solver turned out to be the most efficient and did not run out of memory even at $n=15$. 


\section{Algorithm To Ensure No Extra Tuples}
\label{sec:app:noExtra}
\newstuff{
The presence of additional tuples 
(created, for example, due to repeated relations or foreign key dependencies) 
may change the intended result of a query on the generated test dataset.
For some cases like constrained aggregation and subqueries the additional tuples 
may prevent the generation of desired tuples, and the killing of mutations may be affected.
To avoid the change in the intended result we add constraints preventing  
additional tuples from altering the result; details are described in Algorithm~\ref{algo:noExtraTuples}.
We assume for now that the query tree has only joins and selections.

The algorithm takes as input (1) the query tree for which we do not intend to generate any additional tuples,
(2) the tuples generated for the query tree and 
(3) additional selection conditions for correlation conditions or group by attributes equated to a particular value.
\eat{The algorithm asserts constraints to ensure that for every combination of tuples such that at least one tuple is 
not present in the allowed tuple range, at least one of the selection, 
additional selection or join conditions in the query tree is not true.}

The first step of the algorithm is to create a list of relations, along with the join and selection conditions for the 
given query tree, which we call flattening.
To flatten the tree we recursively traverse the tree.
For INNER JOIN we add both its left and right children to the flattened tree i.e.\ the function makes calls flatten(left) and flatten(right) and returns the union of the lists along with the join conditions.
For the LEFT OUTER JOIN `no extra tuples' can only be ensured if there is no extra tuple from the left input.
We consider only left input for flattening i.e.\ the function calls flatten(left) and returns the list returned by the function.
Similarly for RIGHT OUTER join we consider only the right input for flattening.
For a relation, flattening returns the relation with its selection conditions.
For example, we flatten ($R_1$ $\LOJoin_{\theta 1}$ $R_2$) $\Join_{\theta 2}$ ($R_3$ $\Join_{\theta 3}$ $R_4$) 
to $\Join_{\theta 2,\theta 3}$($R_1,R_3,R_4$).

In the subsequent steps we take the join conditions present in the flattened tree and assert constraints to ensure that 
for every combination of tuples such that at least one tuple is 
not present in the allowed tuple range, at least one of the selection, 
additional selection or join conditions in the flattened query is not satisfied.

\begin{algorithm}
  \renewcommand{\algorithmicrequire}{\textbf{Inputs:}}
    \renewcommand{\algorithmicensure}{\textbf{Output:}}
\begin{algorithmic}[1]
\REQUIRE $T$ = Query tree \\
\hspace{0.7cm}$AT$ = Map of allowed tuples for each relation \\
\hspace{0.7cm}$ASel$ = Additional selection conditions

\ENSURE constraints to ensure no tuple is projected from the subquery 

\STATE $FT$ = flattenTree($T$)
\eat{
\FOR{relation $R$ in $FT$}
  \STATE Let the selection conditions on $R$ be $S_1 \wedge S_2 \wedge ...S_n$
  \STATE Let the additional selection conditions on $R$ be $AS_{R1}$, $AS_{R2}$ AND ...$AS_{Rn}$ 
  \STATE selCond($R$)[i] $=$ $S_1[i]$ AND $S_2[i]$ ... AND $S_n[i]$ AND $AS_{R1}[i]$ AND $AS_{R2}[i]$  .. AND $AS_{Rm}[i]$, where $S_1[i]$ is the constraint for $S_1$ for the $i^{th}$ tuple 
\ENDFOR
  \ELSIF{$R$ is an aggregate}
  \STATE genConstraintsForNoExtraTuples($R$.CHILD)
\ELSIF{$R$ is a LEFT OUTER JOIN}
  \STATE constraints$\leftarrow$genConstraintsForNoExtraTuples($R$.LEFT)
\ELSIF{$R$ is a RIGHT OUTER JOIN}
  \STATE constraints$\leftarrow$genConstraintsForNoExtraTuples($R$.RIGHT)}

  \STATE Let the join conditions for FT be $J_1,J_2..J_c$
  \STATE Let the relations for the join be $R_1, R_2, .. R_m$
  
  \STATE constraints $\leftarrow$ NOT EXISTS $i_{r_1} \in R_1, ... i_{r_n} \in R_n |$  \\ $(i_{r_1}\notin AT[R_1] \vee i_{r_2} \notin AT[R_2] ... \vee i_{r_m} \notin AT[R_m])$ \\ $\wedge (selCond(R_1,i_{r_1}) \wedge selCond(R_2, i_{r_2}) ... \wedge selCond(R_m, i_{r_n})$ $ 
  \wedge$  $J_1' \wedge J_2' ... \wedge J_c')$,\\
  where $J_k'$ is join condition $J_k$ applied to the tuples identified by $i_{r_1}..i_{r_n}$

\RETURN{constraints}
\\
\end{algorithmic}
\textbf{selCond(R,i)}
\begin{algorithmic}[1]
\STATE Let the selection conditions on $R$ be $S_1,S_2,...S_n$
  \STATE Let the additional selection conditions on $R$ be $ASel_{R1}$, $ASel_{R2}$ AND ...$ASel_{Rn}$ 
  \RETURN {$S_1[i]$ AND $S_2[i]$ ... AND $S_n[i]$ AND $ASel_{R1}[i]$ AND $ASel_{R2}[i]$  .. AND $ASel_{Rm}[i]$}, where $S_1[i]$ is the constraint for the selection condition $S_1$ on the $i^{th}$ tuple of $R$
\end{algorithmic}

\caption{ : genConstraintsForNoExtraTuples}
\label{algo:noExtraTuples}
\end{algorithm}

We implement the condition, $i_{r_1} \in AT[R_1]$ by checking that the 
primary key value is not equal to the primary key of any tuple in $AT[R_i]$. 
This is because if a tuple outside the allowed tuples range has the same primary key as a tuple in the allowed tuple range, the tuples are identical.

If the query tree contains GROUP BY attributes and aggregations,
we consider the input to these operators for flattening.
We currently do not handle flatting conditions in subqueries in this algorithm. 

In practice, we unfold the expression in Step 5, to remove the NOT EXISTS quantifier and replace it by conditions for each combination of tuples, in the tuple range, for which we generate data. 
Such unfolding speeds up constraint solving in CVC3 solver as noted earlier in \cite{xdata:icde11}.

}

\section{Completeness}
\label{sec:app:complete}

\newstuff{

Shah et al.\ in \cite{xdata:icde11} present completeness results for join, 
comparison operator and aggregation mutations on a limited space of queries.
In this section, we consider the completeness of our techniques for the wider class of operators
and mutations considered in this paper.

\subsection{Types of Result Difference}
Our techniques for killing mutations generate differences in the result of a mutated operator, 
which may be classified into several types: 
\begin{enumerate}
 \item \emph{Tuple Existence}: A dataset that results in some tuples being present in the result 
 of the original operator, but not in the result of the mutated operator, or vice versa;
 the tuples in one result are a superset of the tuples in the other. 
 
 Tuple existence differences are easy to propagate and relatively easy to generate and are thus
 the preferred type of difference for our data generation techniques.

\emph{Empty Result Difference}: This is a stronger version of tuple existence, where one of the results
is empty while the other is non-empty.  This is needed in case of exists/not-exists subqueries and
a few other cases.
 
\item \emph{Tuple Count}: For some cases like DISTINCT clause it is not feasible to kill mutations
by tuple existence and we generate datasets that produce different numbers of tuples 
(which may be duplicates) for the correct query and the mutation.

\item \emph{Value Difference}: For some other cases like mutations between different aggregate functions,
the above differences cannot be generated, but we instead generate datasets where the correct operator 
and the mutated operator produce different values for one or more attributes.
\end{enumerate}

\subsection{Approach to Showing Completeness}

Our approach to showing completeness for a given class of queries, for a given space of mutations,
is as follows.  The operators we consider are selections, joins, aggregates, projections, subquery, set and GROUP BY. 
Our proof is in terms of relational algebra tree. 
For each possible operator $O_i$, we need to show that: 
\begin{enumerate}
 \item For each non-equivalent mutation, $o_{i}'$ of an occurrence $o_i$ of $O_i$, 
 we generate at least one set of constraints that 
 would result in a difference in the result of $o_i$ compared to $o_{i}'$.
 We describe the possible differences in the result shortly.
 \item For each dataset that is not targeted at mutations of an instance $o_i$ of $O_i$, the constraints generated
   for $o_i$ ``propagate'' certain differences generated in an input of $o_i$ to the result of $o_i$.
   Note that the difference in the output of $o_i$ may not be the same as the difference in the input of $o_i$.
   To ensure completeness in general, every difference should get propagated, but in several cases our
   techniques only propagate some of the differences.

\item For each dataset,
we should add only necessary constraints for data generation and mutation killing.
Removing some constraints could result in a dataset that is not able to kill the intended mutations.
Adding constraints that are not necessary conditions could make the constraints unsatisfiable, even if a solution actually exists for the dataset. 
For most operators, we only generate necessary conditions. 
In some cases, we do not achieve this as discussed in Section~\ref{sec:app:complete:operator}.
\end{enumerate}

Our arguments for completeness are based on the set of constraints we create for data generation.
Note that we use an SMT solver to solve the constraints and generate a dataset, and
SMT solvers are, in general, not complete; however, they have been found to work well in practice.

\subsection{Completeness for Operators Considered}
\label{sec:app:complete:operator}

The operators we consider dataset generation and mutation killing are as follows
\begin{enumerate}
 \item \emph{Selection Operators}
 
\emph{Killing Mutations}:
For mutations in the selection predicates such as  
comparison operator mutations, mutations between conjunctions and disjunctions, string mutations, IS NULL mutations and where clause subquery connective mutations,
mutation killing is ensured by tuple existence as described in Sections~\ref{sec:bg:mutation}, \ref{sec:string}, 
\ref{sec:nulls} and \ref{sec:subquery}. 

It can be seen that only necessary conditions are added to kill the mutations.

\emph{Propagating Difference}:
The same values as input to the selection will be propagated up 
since we assert the selection condition to be true for the tuples 
that are input to the selection as described in \cite{xdata:icde11}. 
Hence irrespective of the result difference technique used to kill the mutation below the selection operator,
the result difference will be propagated up the selection operator. 

We assert only necessary constraints to propagate the mutations.

\item \emph{Join Operators}

\emph{Killing Mutations}: 
\begin{itemize}
 \item \emph{Join Type Mutations}: 
As discussed in \cite{xdata:icde11} (summarized in Section~\ref{sec:bg:mutation}), mutation of INNER JOIN vs.\ any outer join is killed using tuple existence. 
Mutations of LEFT OUTER JOIN vs.\ RIGHT OUTER JOIN are killed by value difference.
Mutations to FULL OUTER JOIN are killed by value difference. 

It can be seen from \cite{xdata:icde11} that only necessary conditions are asserted to kill the join type mutations.

\item \emph{Missing or Additional Join Conditions}: As discussed in Section~\ref{sec:joincond}, mutations of missing or 
 additional join conditions are killed by tuple existence. 
 
 It can be seen from Section~\ref{sec:joincond} that only necessary conditions are asserted to kill these mutations.
\end{itemize}

\emph{Propagating Difference}:
Data generation for joins is done by creating matching tuples for input to the join conditions.
Hence, the same values as input to the join will be propagated up. 
Hence, the result difference below the join will be propagated up for all types of result differences. 
It can be seen from \cite{xdata:icde11} that only necessary conditions are asserted to propagate the differences.


\item \emph{Aggregation Operators}

\emph{Killing Mutations}:
\begin{itemize}
 \item \textit{Aggregation Operator Mutations}: 
As discussed in \cite{xdata:icde11} (summarized in Section~\ref{sec:bg:mutation}) 
aggregation mutations are killed by value difference of result of the aggregate as compared to its mutations.

Only necessary conditions are asserted to kill unconstrained aggregation mutations.
For constrained aggregation, as described in Section~\ref{sec:cagg:dgen} 
we may add constraints that are not necessary for aggregates on join results. 
For aggregation on a single relation, we assert necessary constraints only.

\item \textit{GROUP BY attribute mutations}: 
In case there is no HAVING clause above the GROUP BY attribute, the mutation of changes GROUP BY attributes is killed by tuple count as described in Section~\ref{sec:group}. In case there is a HAVING clause above the GROUP BY, the mutation is killed using tuple existence at the HAVING clause which is also described in Section~\ref{sec:group}. 

Only necessary conditions are asserted for killing these mutations.

\end{itemize}

\emph{Propagating Difference}:
Not all differences due to mutations below are propagated by aggregation operators.

\begin{itemize}
 \item First consider the aggregates SUM (DISTINCT), AVG (DISTINCT), MIN and MAX.  
If the mutation below the aggregate is killed by tuple existence then the aggregate 
produces a zero result for the case where the tuples exists and a non-zero result 
for the cases where the tuple does not exist (aggregated attributes are asserted to be non-zero). 
Hence, a mutation that is killed by tuple existence below will result in value difference at the aggregate. 
For mutations below that are killed tuple count or by value difference, 
the aggregate may produce the same result and the mutation might not get killed.
Mutations below killed by value difference will produce a value difference at the aggregate if 
the value difference  at the mutated node is a NULL vs.\ NOT NULL difference. 

\item 
Now consider the aggregates  COUNT and COUNT DISTINCT. Mutations killed by tuple existence will be killed by COUNT or COUNT DISTINCT. Mutations killed by tuple count will produce a value difference for COUNT and hence the difference will  be propagated. 
For COUNT DISTINCT, mutations killed by tuple count may produce the same values and hence may not get propagated.
Mutations below killed by value difference will produce a value difference in COUNT or COUNT DISTINCT only 
if the value difference at the mutated node is a NULL vs.\ NOT NULL difference. 
\end{itemize}

For unconstrained aggregation, no constraints are added for data generation. 
The constraints added for mutation killing are necessary as described in Section~\ref{sec:bg:mutation}. 
For constrained aggregation, as described in Section~\ref{sec:cagg:dgen} 
we may add constraints that are not necessary.
For aggregation on a single relation, we do not face this issue and hence only necessary constraints are added.

\item \emph{Projection Operator (non-duplicate removing)}

\emph{Killing Mutations}: Currently we do not target mutations in projections. 
Mutations due to adding or removing attributes would get caught if present at the top of the query tree.
We currently do not generate any datasets to catch projection mutations.
Killing projection mutations could be done by asserting attributes in the projection list have different values wherever possible, an area of future work.

\emph{Propagating Difference}:
A mutation below that is killed by tuple existence or by tuple count will be preserved after projection. 
A value difference will be propagated up only if the attribute whose value is different is present in 
the projected attributes.

We do not add any constraints for projection and hence trivially only necessary constraints are added.

\item \emph{DISTINCT operator}

\emph{Killing Mutations}: DISTINCT clause mutation is killed by tuple count as shown in Section~\ref{sec:distinct}. 

Only necessary constraints are asserted to kill DISTINCT clause mutations as shown in Section~\ref{sec:distinct}.

\emph{Propagating Difference}:
If a mutation below the distinct clause is killed by tuple existence or 
by value difference, then the DISTINCT clause will also preserve the respective property. 
However if the mutation below the DISTINCT clause is killed by tuple count
the DISTINCT clause may not be able to preserve the difference.

We do not add any constraints for the DISTINCT clause and hence trivially only necessary constraints are added.

\item \emph{Subquery Operator}

\emph{Killing Mutations}:
Mutations of the subquery connectives (EXISTS, NOT EXISTS, IN, NOT IN, ALL, ANY and scalar subqueries) in the WHERE clause comes under selection mutation and is discussed earlier in the bullet on selection mutation.
We currently only handle scalar subqueries of the form \texttt{SSQ \relop\ attr/value}, where \texttt{SSQ} is a scalar subquery, \texttt{attr} is an attribute from the outer block of query and \texttt{value} is a constant.

For subqueries other than scalar subqueries with aggregation, only necessary constraints for killing mutations are asserted.
Scalar subqueries with aggregation use constrained aggregation techniques and hence constraints that are not necessary may also be added if the subquery contains more than one relation.

\emph{Propagating Difference}:
For mutations of operators in the subquery, the subquery connective preserves tuple existence by ensuring empty result difference. 
Other types of mutation killing may not be propagated up; these mutations are equivalent in many but not all cases.
Refer Section~\ref{sec:subq:mutation} for details. 

Only necessary conditions for propagating differences are asserted as can be seen from Section~\ref{sec:subq:mutation}.

\item \emph{Set Operators}

\emph{Killing Mutations}:
Mutations killing for UNION vs UNION ALL, INTERSECT vs.\ INTERSECT ALL and 
EXCEPT vs.\ EXCEPT ALL is done by tuple count.
Other mutations are killed by tuple existence as shown in Section~\ref{sec:set:set_mutation}.

Only necessary constraints required for killing set operator mutations are asserted as can be seen from Section~\ref{sec:set:set_mutation}.

\emph{Propagating Difference}:
As explained in Section~\ref{sec:set:mutation} mutations below the set operator 
are propagated up for all mutations.

Only necessary conditions for propagating differences are asserted as can be seen from Section~\ref{sec:set:mutation}.

\end{enumerate}

\subsection{Summary}
Our data generation techniques are complete for  killing a given mutation on a given operator of a given query tree if 
\begin{enumerate}
 \item The constraint generation technique creates a difference at the mutated operator
 \item The difference at the mutated node is propagated up the query tree to the root i.e. each ancestor node 
 propagates the difference type generated by its child on the path from the mutated node.
 \item Only necessary constraints are added for data generation.
\end{enumerate}
If the above properties are satisfied for all operators in the query and all mutations of the operator in a space of mutations, then our data generation techniques are complete for the query under the space of mutations considered. 

Although not complete, in practice our data generation techniques work well. 
Our experimental results in Section~\ref{sec:expt} show that we are able to generate test data and kill mutations for a large variety of common queries.  
}

\section{Test Cases and Results for Experiments}
\label{sec:app:expt}
\reminder{check section name}
In this section, we list \newstuff{results of the grading tool} and the test cases that were used for the experiments described in Section~\ref{sec:expt}.

\subsection{Grading Tool Results}

Result of the grading tool experiment is listed in Table~\ref{tab:query:eval}.

\begin{table}
\begin{center}
\begin{small}
\tabcolsep=0.15cm
\begin{tabular}{|c|c|c|c|c|c|c:c|}
    \hline
    \multirow{2}{*}{\textbf{QId}}&\multirow{1}{*}{\textbf{Que-}}& 
\textbf{XData}&
\textbf{USm}&
\textbf{ULg}&
\textbf{TA}&
\multicolumn{2}{c|}{\textbf{Plan}} \\
    \cline{3-8}
    &\multicolumn{1}{c|}{\textbf{ries}}&\textbf{$\times$} & {$\times$} & {$\times$}   &{$\times$}  & {$\surd$}&{$?$}\\        
\hline
CQ1	& 55	& 2	& 2	& 2		& 2	&51	&4\\
\hline
CQ2	& 57	& 1	& 1	& 1		& 1	&54	&3\\
\hline
CQ3	& 71	& \textbf{13}	& 12		&1	& 1	&3	&68\\
\hline
CQ4	& 78	& \textbf{26}	& \textbf{26}	& 3		& 1	&52	&26\\
\hline
CQ5	& 72	& \textbf{23}	& 11	& 16		& 13	&43	&29\\
\hline
CQ6	& 61	& 6	& 6	& 6		& 2	&55	&6\\
\hline
CQ7	& 77	& \textbf{25}	& 23	& 3		& 24	&3	&74\\
\hline
CQ8	& 79	& \textbf{33}	& 12	& 14		& 16	&2	&77\\
\hline
CQ9a	& 80	& 68	& 24	& 70		& 23	&2	&78\\
CQ9b	& 80	& 71	& 24	& 70		& 23	&3	&77\\ \hdashline

CQ9	& 80	& \textbf{72}	& 24	& 70		& 23	&5	&75\\
\hline
CQ10	& 74	& 1	& 1	& 1		& 0	&34	&40\\
\hline
CQ11	& 69	& 16	& 16	& 16		& 16	&51	&18\\
\hline
CQ12	& 70	& \textbf{8}	& 3	& 7		& 7	&38	&32\\
\hline
CQ13	& 72	& 9	& 9	& 9		& 7	&3	&69\\
\hline
CQ14	& 67	& \textbf{34}	& 14	& 10		& 32	&2	&65\\
\hline

  \end{tabular}
\end{small}
\caption{Query grading results}
\label{tab:query:eval}
\end{center}
\end{table}

\newstuff{
The column labeled Queries lists the number of student queries that were submitted.
Columns labeled XData, USm, ULg and TA show the number of incorrect queries caught by these techniques.
Plan gives the number of queries labeled as correct and ones for which the plan is not able to determine correctness.
Wherever our technique and/or some of the datasets find more incorrect queries than
others, we have highlighted the results in bold.
}

\subsection{Test Queries for Constrained Aggregation}
For the experiment involving constrained aggregation, we used the following set of queries:

\begin{small}
\begin{enumerate}[{CA}1:] \itemsep=0.6em
 \query{SELECT c.dept\_name, SUM(c.credits) \\
 FROM course c INNER JOIN department d \\
 \qindent ON (c.dept\_name = d.dept\_name)  \\ 
 GROUP BY c.dept\_name \\HAVING SUM(c.credits)$>$10
 AND COUNT(c.credits)>1}
 
 \query{SELECT c.dept\_name, SUM(i.salary) \\
 FROM course c INNER JOIN department d\\ \qindent ON (c.dept\_name = d.dept\_name) \\
 \qindent INNER JOIN instructor i \\ \qindent ON (d.dept\_name = i.dept\_name) \\
 GROUP BY c.dept\_name\\ HAVING SUM(i.salary)$>$100000 \\ \qindent AND MAX(i.salary)$<$75000}
 
 \query{SELECT c.dept\_name, SUM(d.budget) \\
 FROM course c INNER JOIN department d \\ \qindent ON (c.dept\_name = d.dept\_name) \\
 \qindent INNER JOIN teaches t \\\qindent ON (c.course\_id = t.course\_id) \\
 GROUP BY c.dept\_name \\HAVING SUM(d.budget)$>$100000 AND COUNT(d.budget)$>$1}
 
 \query{SELECT c.dept\_name, AVG(i.salary) \\
 FROM course c INNER JOIN department d \\ \qindent ON (c.dept\_name = d.dept\_name) \\
 \qindent INNER JOIN teaches t \\ \qindent ON (c.course\_id = t.course\_id) \\
 \qindent INNER JOIN instructor i \\ \qindent ON (d.dept\_name = i.dept\_name) \\
 GROUP BY c.dept\_name \\
 HAVING AVG(i.salary)$>$50000 AND COUNT(i.salary)=3}
 
 \query{SELECT t.semester, SUM(c.credits) \\FROM department d INNER JOIN teaches t  \\ \qindent ON (d.budget = t.year + 4)  \\ \qindent INNER JOIN course c \\ \qindent ON (c.dept\_name = d.dept\_name)  \\GROUP BY t.semester \\HAVING AVG(c.credits)$>$2 AND COUNT(d.building)=2} 

  \query{SELECT id \\FROM course NATURAL JOIN department \\ \qindent NATURAL JOIN student NATURAL JOIN takes \\ \qindent NATURAL JOIN section\\   GROUP BY id,dept\_name HAVING COUNT(dept\_name)$>$1}

  \query{SELECT distinct dept\_name \\FROM course WHERE credits =\\ \qindent (SELECT MAX(credits) \\ \qindent FROM course NATURAL JOIN department \\ \qindent WHERE  title=`CS' \\ \qindent GROUP BY dept\_name HAVING COUNT(course\_id)$>$2)}

  \query{SELECT id,name FROM \\ \qindent (SELECT id,time\_slot\_id,year,semester \\ \qindent FROM takes NATURAL JOIN section \\ \qindent GROUP BY id,time\_slot\_id,year,semester \\ \qindent HAVING COUNT(time\_slot\_id)$>$1) \\as s NATURAL JOIN student \\GROUP BY id, name \\HAVING COUNT(id)$>$1}

  \query{SELECT SUM(T) as su FROM \\ \qindent (SELECT year as T \\ \qindent FROM teaches  NATURAL JOIN instructor \\ \qindent GROUP BY year, course\_id HAVING COUNT(id)$>$4) \\as temp GROUP BY T}

\end{enumerate}
\end{small}

\subsection{Test Queries for Subquery}
For the experiment involving subqueries, we used the following set of queries:

\begin{small}
\begin{enumerate}[{SQ}1:] \itemsep=0.6em
\query{SELECT * FROM department  d \\WHERE d.dept\_name IN (SELECT c.dept\_name \\ \qindent FROM course c WHERE c.credits $>$ 2)  }

\query{SELECT * FROM course  c \\WHERE EXISTS (SELECT * FROM department d \\ \qindent WHERE c.dept\_name = d.dept\_name)}    
    
\query{SELECT * FROM takes t \\WHERE NOT EXISTS (SELECT * FROM section \\ \qindent WHERE t.year=section.year AND year = 2010)}    
    
\query{SELECT * FROM course c \\WHERE credits $>$ 3 AND \\EXISTS (SELECT * FROM department d \\ \qindent WHERE d.dept\_name = c.dept\_name)}     
    
\query{SELECT course\_id, title \\FROM course NATURAL JOIN section \\WHERE SEMESTER = `Spring' AND year = 2010 AND \\course\_id IN (SELECT course\_id FROM prereq \\ \qindent WHERE prereq\_id = `CS-201')}

\query{SELECT course\_id, TITLE \\FROM course NATURAL JOIN section \\WHERE SEMESTER = `Spring' AND year = 2010 AND \\\qindent course\_id NOT IN (SELECT course\_id FROM prereq \\ \qindent \qindent WHERE prereq\_id = `CS-201') }

\query{SELECT name FROM instructor \\WHERE salary $>$ALL (SELECT salary \\ \qindent FROM instructor WHERE dept\_name = `Biology') }

\query{SELECT name FROM instructor \\WHERE salary $>$ (SELECT AVG(salary) \\ \qindent FROM instructor WHERE dept\_name = `Physics')}

\query{SELECT * FROM student \\WHERE tot\_cred $>$ (SELECT SUM(credits) \\ \qindent FROM takes INNER JOIN course USING(course\_id) \\ \qindent WHERE student.id=takes.id)} 
 
\query{SELECT * FROM student \\WHERE tot\_cred $<$ALL (SELECT SUM(credits) \\ \qindent FROM takes INNER JOIN course USING(course\_id) \\ \qindent WHERE dept\_name=`History')} 

\end{enumerate}
\end{small}

\subsection{Correct Queries for Grading Tool}
Following are the correct queries that were used in the experiment to grade student queries:

\begin{small}
\begin{enumerate}[{CQ}1:] \itemsep=0.6em

\query{SELECT course\_id, title FROM course}

\query{ SELECT course\_id, title FROM course \\WHERE dept\_name$=$ `Comp. Sci.'  }

\query{ SELECT DISTINCT course\_id, title, id\\FROM course NATURAL JOIN teaches \\WHERE teaches.semester $=$ `Spring' \\ \qindent AND teaches.year $=$ `2010' }

\query{SELECT DISTINCT student.id, student.name \\FROM takes NATURAL JOIN student \\WHERE course\_id $=$`CS-101'  }

\query{SELECT DISTINCT course.dept\_name \\FROM course NATURAL JOIN section \\WHERE section.semester $=$ `Spring' \\ \qindent AND section.year $=$ `2010' }

\query{SELECT course\_id, title FROM course \\WHERE credits $>$ 3  }

\query{ SELECT course\_id, COUNT(DISTINCT id) \\FROM course NATURAL LEFT OUTER JOIN takes \\GROUP BY course\_id }

\query{ SELECT DISTINCT course\_id, title \\FROM course NATURAL JOIN section \\WHERE semester $=$ `Spring' AND year $=$ 2010 AND \\\qindent course\_id NOT IN (SELECT course\_id FROM prereq) }

\item
\begin{enumerate}[a)]\itemsep=0.3em
 \query{ WITH s as \\\qindent (SELECT id, time\_slot\_id, year, semester \\\qindent FROM takes NATURAL JOIN section \\ \qindent GROUP BY id, time\_slot\_id, year, semester \\ \qindent HAVING COUNT(time\_slot\_id)$>$1)
     \\SELECT DISTINCT id,name \\FROM s NATURAL JOIN student }

\query{SELECT DISTINCT A.id, A.name FROM \\\qindent (SELECT * FROM student NATURAL JOIN takes \\\qindent \qindent \qindent NATURAL JOIN section) A, \\\qindent (SELECT * from student NATURAL JOIN takes \\\qindent \qindent \qindent NATURAL JOIN section) B \\WHERE A.name = B.name AND A.year = B.year  \\\qindent AND A.course\_id $<>$ B.course\_id \\\qindent AND A.semester = B.semester \\\qindent AND A.time\_slot\_id = B.time\_slot\_id  }
\end{enumerate}
\query{SELECT DISTINCT dept\_name FROM course \\WHERE credits$=$(SELECT MAX(credits) FROM course)  }

\query{ SELECT DISTINCT instructor.id, name, course\_id \\FROM instructor LEFT OUTER JOIN TEACHES \\ \qindent ON instructor.id $=$ teaches.id  }

\query{ SELECT student.id, student.name FROM student \\WHERE lower(student.name) like `\%sr\%' }

\query{SELECT id,name FROM student s WHERE \\NOT EXISTS \\ \qindent(SELECT * FROM student t 
NATURAL JOIN takes \\ \qindent WHERE s.id=t.id AND takes.year=2010 \\ \qindent AND 
takes.semester=`Spring')  }

\query{SELECT DISTINCT * FROM takes t \\WHERE \\\qindent (NOT EXISTS (SELECT id,course\_id \\ \qindent \qindent FROM takes s \\\qindent \qindent WHERE grade $!=$ `F' AND t.id $=$ s.id \\\qindent \qindent \qindent AND t.course\_id$=$s.course\_id) \\\qindent \qindent \qindent AND t.grade IS NOT NULL) \\\qindent OR (t.grade $!=$ `F' AND t.grade IS NOT NULL)  }

\end{enumerate}
\end{small}

\balance
\end{document}